\documentclass[prd,superscriptaddress,twocolumn,nofootinbib]{revtex4}
\usepackage{amsfonts,amsmath,amssymb}
\usepackage{graphicx}% Include figure files
\usepackage{dcolumn}% Align table columns on decimal point
\usepackage{bm}% bold math
\newcommand{\B}[1]{{\mathbb #1}}
\newcommand{\C}[1]{{\mathcal #1}}

\newcommand{\bea}{\begin{eqnarray}}
\newcommand{\eea}{\end{eqnarray}}
\newcommand{\beq}{\begin{equation}}
\newcommand{\eeq}{\end{equation}}
\newcommand{\nn}{\nonumber}

\newcommand{\half}{\frac 12}
\newcommand{\third}{\frac 13}

\newcommand{\Slash}[1]{{\ooalign{\hfil#1\hfil\crcr\raise.167ex\hbox{/}}}}

\begin{document}

%%%%%%%%%%%%%%%%%%%%%%%%%%%%%%%%%%%%%%%%%%%%%%%%

\preprint{arXiv:yymm.nnnn}

%%%%%%%%%%%%%%%%%%%%%%%%%%%%%%%%%%%%%%%%%%%%%%%%

\title{Higgs-lepton inflation in the supersymmetric minimal seesaw model
}

\author{Masato Arai}
\email{masato.arai(AT)fukushima-nct.ac.jp}
\affiliation{
Fukushima National College of Technology,
%30 Aza-Nagao, Tairakamiarakawa, 
Iwaki, 
Fukushima 
970-8034, Japan}
%\affiliation{
%IEAP, Czech Technical University in Prague, Horsk\' a 3a/22, 128 00 Prague 2, Czech Republic}
\author{Shinsuke Kawai}
\email{kawai(AT)skku.edu}
\affiliation{
Institute for the Early Universe (IEU),
11-1 Daehyun-dong, Seodaemun-gu, Seoul 120-750, Korea} 
\affiliation{Department of Physics, 
Sungkyunkwan University,
Suwon 440-746, Korea}
\author{Nobuchika Okada}
\email{okadan(AT)ua.edu}
\affiliation{
Department of Physics and Astronomy, 
University of Alabama, 
Tuscaloosa, AL35487, USA} 

\date{\today}% It is always \today, today,

%%%%%%%%%%%%%%%%%%%%%%%%%%%%%%%%%%%%%%%%%%%%%%%%
%%%%% Abstract %%%%%
\begin{abstract}
We investigate a scenario of cosmological inflation realised along a flat direction of the
minimal seesaw model embedded in supergravity with a noncanonical R-parity violating 
K\"{a}hler potential.
It is shown that with appropriate seesaw parameters the model is consistent 
with the present observation of the cosmological microwave background (CMB) as well as
with the neutrino oscillation data.
It is also shown that the baryon asymmetry of the Universe can be generated through leptogenesis.
The model favours supersymmetry breaking with the gravitino as the lightest superparticle, and
thus indicates the gravitino dark matter scenario.
An interesting feature of this model is that the seesaw parameters are constrained by the CMB spectra.
The $2$-$\sigma$ constraints from the 9-year WMAP data yield a mild lower bound on the 
seesaw mass scale $\gtrsim$ TeV.
We expect that the observation by the Planck satellite will soon provide more stringent constraints.
%and the forthcoming CMB polarisation experiments 
%such as POLARBEAR and LiteBIRD,
The phenomenological and cosmological implications of the R-parity violation 
are also discussed.
\end{abstract}

%%%%%%%%%%%%%%%%%%%%%%%%%%%%%%%%%%%%%%%%%%%%%%%%

\maketitle

%%%%%%%%%%%%%%%%%%%%%%%%%%%%%%%%%%%%%%%%%%%%%%%%
%%%%% Body of the Paper %%%%%
%%%%%%%%%%%%%%%%%%%%%%%%%%%%%%%%%%%%%%%%%%%%%%%%

\section{Introduction}
In cosmology, the particle physics origin of inflation remains as an unsolved problem.
In search of a realistic inflationary model one may take two possible approaches.
One is the top-down approach, starting with a unifying theory including gravity.
String cosmology \cite{McAllister:2007bg}, loop quantum cosmology \cite{Ashtekar:2011ni}, and 
cosmological model building based on F-theory grand unification \cite{Heckman:2008jy}
fall into this category.
The top-down approach is an ambitious program to build a consistent scenario from the 
first principles. 
The other is the bottom-up approach, that is, to investigate embedding of inflation into particle
physics models that are confirmed at low energies, starting from the Standard Model (SM) and
its various extensions. 
The bottom-up approach is advantageous in predictability and falsifiability. 
In view of the remarkable progress and ever-increasing precision in observational 
cosmology \cite{Bennett:2012fp,Hinshaw:2012fq}
%,{Komatsu:2010fb}, 
one could hope that in the near future observation will be able to narrow
SM-based inflationary models down to a handful of candidates. 
In this paper we take the bottom-up approach and propose a model of inflation, which we believe
to be a promising and testable candidate.

Recently, there has been a revival of interest in cosmological inflation within the SM,
where the Higgs field, nonminimally coupled to gravity, is identified as an inflaton 
\cite{CervantesCota:1995tz,Bezrukov:2007ep,Barvinsky:2008ia,DeSimone:2008ei,Bezrukov:2009db,Bezrukov:2010jz,Barvinsky:2009fy,Barvinsky:2009ii,Barbon:2009ya,Burgess:2009ea,Burgess:2010zq,Hertzberg:2010dc,Lerner:2009na,Lerner:2010mq}
(see also \cite{Starobinsky:1980te,Spokoiny:1984bd,Futamase:1987ua,Salopek:1988qh,Makino:1991sg}
for older proposals of nonminimally coupled inflation models).
Interestingly, it requires the Higgs mass to be $126 - 194$ GeV, consistent with the mass scale
suggested by the Large Hadron Collider experiments \cite{:2012gk,:2012gu}.
The Higgs inflation model also predicts small tensor-mode perturbation that fits remarkably well 
with the present-day cosmological microwave background (CMB) observation.
The success of the model comes at the cost of an extremely large coupling $\xi\sim10^4$ between
the Higgs field and the background curvature, which could lead to the violation of the unitarity bound
\cite{Barbon:2009ya,Burgess:2009ea,Burgess:2010zq,Lerner:2009na,Lerner:2010mq,Hertzberg:2010dc} (for the controversy see also \cite{Barvinsky:2009fy,Barvinsky:2009ii,Bezrukov:2010jz,Ferrara:2010yw,Ferrara:2010in}). 
Also, the model suffers from the usual hierarchy problem of the SM, namely
the Higgs potential is destabilised due to radiative corrections
\cite{Barvinsky:2008ia,DeSimone:2008ei,Barvinsky:2009fy,Barvinsky:2009ii}.
As is well known, the hierarchy problem is solved, or at least tamed, by introducing
supersymmetry.
Supersymmetric versions of the Higgs inflation model were considered in 
\cite{Einhorn:2009bh,Ferrara:2010yw,Ferrara:2010in},
and it was shown that while embedding into the minimal supersymmetric Standard Model (MSSM) 
is not possible (as the field content is too restrictive), the next-to-minimal supersymmetric SM 
successfully accommodates Higgs inflation.
Embedding into supersymmetric grand unified theory \cite{Arai:2011nq} 
(see also \cite{Einhorn:2012ih}) and other supersymmetric 
SMs \cite{Pallis:2011ps} have also been shown to be possible.

Strictly speaking, the SM is not entirely a satisfactory model of particle phenomenology any longer, 
even apart from the hierarchy problem:
it fails to explain neutrino oscillations; it does not contain good candidates for the dark matter;
the generation of the baryon asymmetry in the electroweak phase transition is known to be
problematic.
Among the simplest extensions of the SM that can solve all these problems is the 
supersymmetric extension of the seesaw model \cite{seesaw}.
In \cite{Arai:2011aa} we proposed a new scenario of inflation within the supersymmetric seesaw 
model, inspired by the developments of the nonminimally coupled Higgs inflation models.
Notably, in that scenario the problem associated with the large nonminimal coupling is alleviated.
The purpose of the present paper is to discuss the viability of this inflationary scenario using
realistic neutrino mass parameters. 
We follow the minimalistic guiding principle of Higgs inflation and consider the {\em minimal} 
seesaw model \cite{Frampton:2002qc}, i.e., with two families of the right-handed neutrinos.
There are several other inflationary scenarios based on the supersymmetric seesaw model
\cite{Murayama:1992ua,Murayama:1993xu,Ellis:2003sq,Allahverdi:2006iq,Allahverdi:2006we,Allahverdi:2006cx,BuenoSanchez:2006xk}.
We shall compare the cosmological predictions of these models with ours, and argue that
at least some of these models are soon to be excluded by CMB observations.

The main question we shall address below is whether or not the model allows parameter 
regions that are consistent with the observed baryon asymmetry, the spectrum of the CMB and the
neutrino oscillation data.
The ratio of the baryon density $n_B$ to the entropy density $s$ is
measured to be \cite{Bennett:2012fp,Hinshaw:2012fq}
\beq
Y_B\equiv \frac{n_B}{s}=8.7\times 10^{-11}.
\label{eqn:YB}
\eeq 
As mentioned above, the electroweak baryogenesis within the SM is problematic as it
requires rather implausible strongly first order electroweak phase transition.
We show that in our scenario the thermal leptogenesis \cite{Fukugita:1986hr} with the 
reheating temperature $T_{\rm RH}\approx 10^5$ GeV, combined with the resonant enhancement 
effects \cite{Flanz:1996fb,Pilaftsis:1997jf,Pilaftsis:2003gt} yields the observed amount of the baryon 
asymmetry (\ref{eqn:YB}).
An interesting feature of our scenario is that the CMB spectrum is related to the seesaw mass scale
\cite{Arai:2011aa}.
We investigate this relation in the supersymmetric minimal seesaw model, and show that for a wide 
range of parameters our scenario is consistent with the present CMB observation.
In the near future the CMB data will further constrain the seesaw mass scale.
In our scenario the R-parity needs to be broken.
We discuss that this requirement leads to a scenario of gravitino dark matter.

The paper is organised as follows. 
In the next section we describe the supersymmetric minimal seesaw model on which our
scenario is based.
In Section III the dynamics of inflation is discussed, and in Section IV the reheating process and 
baryogenesis are studied.
The relation between the cosmological parameters and the neutrino mass parameters is 
discussed in Section V. 
Section VI deals with the R-parity violation.
We conclude in Section VII with comments.
Technicalities in computing the baryon asymmetry are relegated to the Appendix.

%%%%%%%%%%%%%%%%%%%%%%%%%%%%%%%%%%%%%%%%%%%%%%%%
\section{The supersymmetric minimal seesaw model}
%%%%%%%%%%%%%%%%%%%%%%%%%%%%%%%%%%%%%%%%%%%%%%%%
Our model is based on the supersymmetric seesaw model, with the superpotential, 
\bea
W&=&
\mu H_uH_d+y_u^{ij} u^c_i Q_j H_u+y_d^{ij} d^c_i Q_j H_d\nn\\
&&+y_e^{ij} e^c_i L_j H_d+y_D^{mi} N_m^c L_i H_u+\half M_m N_m^c N_m^c.
\label{eqn:W}
\eea
Here, $Q$, $u^c$, $d^c$, $L$, $e^c$, $H_u$, $H_d$ are the MSSM superfields, 
$N_m^c$ the right-handed neutrino superfields (having odd $R$-parity), 
$M_{m}$ the corresponding right-handed neutrino mass parameters,
$\mu$ the MSSM $\mu$-parameter and
%The superscripts $c$ stand for the complex conjugation, 
$y_D^{mj}$, $y_u^{ij}$, $y_d^{ij}$, $y_e^{ij}$ are the Yukawa couplings.
The superfields $Q$, $L$, $H_u$, $H_d$ are $SU(2)$ doublets and the contraction using 
the $SU(2)$ invariant 
$i\sigma_2= \left({\begin{array}{cc} 0 & 1 \\ -1 & 0\end{array}}\right)$
%$i\sigma_2= \left({\tiny\begin{array}{cc} 0 & 1 \\ -1 & 0\end{array}}\right)$ 
is implicit, whereas $u^c$, $d^c$, $e^c$ are $SU(2)$ singlets.
In this paper we shall focus on the simplest realistic seesaw model with two right-handed neutrinos
(the minimal seesaw model \cite{Frampton:2002qc}).
Thus the family indices run $m,n,\cdots=1, 2$ for the right-handed neutrinos and $i,j,\cdots =1,2,3$
for the other lepton and the quark superfields.

Using the stationarity condition
$\delta W/\delta N_m^c=y_D^{mi} L_i H_u + M_m N_m^c=0$, 
the superpotential along the flat direction reads
\bea
W_{\rm eff}
&=&N_m^c (y_D^{mi} L_i H_u + M_m N_m^c) -\half M_m N_m^c N_m^c\nn\\
&=&-\half M_m^{-1}
\left(%\frac{1}{M_m} 
y_D^{mi} L_i H_u\right)
\left(%\frac{1}{M_m}
y_D^{mj}L_j H_u\right).
\label{eqn:Weff}
\eea
%We use the unitary gauge for the Higgs field,
The Higgs field develops the vacuum expectation value at low energies,
\beq
\langle H_u\rangle=\left(\begin{array}{c} 0\\ \langle H_u^0\rangle\end{array}\right),
\eeq
where
$\langle H_u^0\rangle =\frac{v}{\sqrt 2} \sin\beta$ with
$v=246$ GeV.
We use $\tan\beta=10$ throughout this paper.
The mass matrix of the left-handed neutrinos obtained from (\ref{eqn:Weff}) is
\beq
m_\nu= m_D^T M^{-1} m_D,
\label{eqn:seesaw}
\eeq
where
$m_D=y_D\langle H_u^0\rangle$
and
$M=\mbox{diag} (M_1, M_2)$.
This is the celebrated seesaw relation.
The Maki-Nakagawa-Sakata (MNS) lepton flavour mixing matrix is parametrized as
\begin{widetext}
\beq
U_{\rm MNS}
=
\left(\begin{array}{ccc}c_{12}c_{13} & s_{12}c_{13} & s_{13}e^{-i\delta} \\-s_{12}c_{23}-c_{12}s_{23}s_{13}e^{i\delta} & c_{12}c_{23}-s_{12}s_{23}s_{13}e^{i\delta} & s_{23}c_{13} \\s_{12}s_{23}-c_{12}c_{23}s_{13}e^{i\delta} & -c_{12}s_{23}-s_{12}c_{23}s_{13}e^{i\delta} & c_{23}c_{13}\end{array}\right)
\left(\begin{array}{ccc}1 & 0 & 0 \\0 & e^{i\sigma} & 0 \\0 & 0 & 1\end{array}\right),
\eeq
\end{widetext}
where
$s_{ij}=\sin\theta_{ij}$, $c_{ij}=\cos\theta_{ij}$, and
$\delta$, $\sigma$ are the CP-violating Dirac and Majorana phases.
The neutrino mass matrix is diagonalised by the MNS matrix %$U_{\rm MNS}$, 
as
\beq
m_\nu=U_{\rm MNS}^* D_\nu U_{\rm MNS}^\dag,
\eeq
where
$D_\nu\equiv\mbox{diag}(m_1, m_2, m_3)$.
% Neutrino oscillation
We use the neutrino mass parameters from the oscillation data 
\cite{Nakamura:2010zzi,An:2012eh},
\bea
&&\sin^2 2\theta_{12}= 0.87,\quad
\sin^2 2\theta_{23}= 1.0, \quad
\sin^2 2\theta_{13}= 0.092,\nn\\
&&\Delta m_{12}^2\equiv m_2^2-m_1^2=7.59\times 10^{-5} \mbox{eV}^2,\nn\\
&&\Delta m_{23}^2\equiv \vert m_3^2-m_2^2\vert = 2.43\times 10^{-3} \mbox{eV}^2.
\label{eqn:oscidata}
\eea
In terms of these the neutrino masses are
\beq
m_1=0,\,
m_2=\sqrt{\Delta m_{12}^2},\,
m_3=\sqrt{\Delta m_{12}^2+\Delta m_{23}^2},
\eeq
for the normal mass hierarchy (NH) and
\beq
m_1=\sqrt{\Delta m_{23}^2-\Delta m_{12}^2},\,
m_2=\sqrt{\Delta m_{23}^2},\,
m_3=0,
\eeq
for the inverted mass hierarchy (IH).
The neutrino mass matrix can be conveniently parametrized as \cite{Casas:2001sr,Ibarra:2003up}
\beq
m_D = \sqrt{M} R\sqrt{D_\nu} U_{\rm MNS}^\dag,
\label{eqn:CIR}
\eeq
where $\sqrt{M}\equiv \mbox{diag}(\sqrt{M_1},\sqrt{M_2})$ and
\beq
\sqrt{D_\nu} =\left\{
\begin{array}{ccc}
\sqrt{D_\nu^{\rm NH}} &=&\left(\begin{array}{ccc}0 & \sqrt{m_2} & 0 \\0 & 0 & \sqrt{m_3}\end{array}\right),\\
\sqrt{D_\nu^{\rm IH}} &=&\left(\begin{array}{ccc}\sqrt{m_1} & 0 & 0 \\0 & \sqrt{m_2} & 0\end{array}\right).
\end{array}\right.
\eeq
In (\ref{eqn:CIR}), $R$ is a $2\times 2$ orthogonal matrix
\beq
R=\left(\begin{array}{cc}\cos w & \sin w \\-\sin w & \cos w\end{array}\right),\quad
w\in{\B C}.
\label{eqn:R}
\eeq
We shall write the real and imaginary parts of $w$ as $w=a+ib$, $a, b\in{\B R}$. 
A merit of the minimal seesaw model is its strong predictive power.
Note that the mass matrix $m_D$ (or equivalently the Dirac Yukawa coupling $y_D$) contains 9 real 
degrees of freedom (6 complex minus 3 phases), 
5 of which are constrained by the oscillation data, namely the two neutrino masses and
the three angles in (\ref{eqn:oscidata}).
In addition, there are one Dirac and one Majorana phases, and the remaining two degrees of 
freedom correspond to the choice of $a$ and $b$.
The right-handed neutrino masses $M_1$, $M_2$ and the Dirac and Majorana phases are not fixed 
by the present experiments.
In our scenario these parameters are subject to the constraints from the CMB, as discussed below.

%%%%%%%%%%%%%%%%%%%%%%%%%%%%%%%%%%%%%%%%%%%%%%%%
\section{Inflationary dynamics and cosmological parameters}
%%%%%%%%%%%%%%%%%%%%%%%%%%%%%%%%%%%%%%%%%%%%%%%%

In this section we describe the construction of the inflation model and discuss its prediction on 
cosmological parameters.
The model, which is a multi-family generalisation of the one introduced in \cite{Arai:2011aa},
has some similarity to the supersymmetric Higgs inflation models 
\cite{Einhorn:2009bh,Ferrara:2010yw,Ferrara:2010in,Arai:2011nq}.
A notable difference from these models is that the nonminimal coupling $\xi$ is allowed to take 
small values and the unitarity violation problem is thus alleviated.

%%%%%%%%%%%%%%%%%%%%%%%%%%%%%%%%%%%%%%%%%%%%%%%%
\subsection{The model of inflation}
We assume that inflation takes place along one of the $L$-$H_u$ D-flat directions and
%The fields $Q$, $u^c$, $d^c$, $e^c$, $H_d$ will be ignored below, as they do not play any 
%r\^{o}le in the physics we discuss.
%
as an initial condition one of the three (scalar components of the) $L_i$ fields,
call it $L_k$ with $k$ fixed, has a large expectation value and dominates the inflationary dynamics
over the other $L_i$'s.
The D-flat direction along $L_k$-$H_u$ can be parametrized as
\beq
L_k=\frac{1}{\sqrt 2}\left(\begin{array}{c} \varphi\\ 0 \end{array}\right),
\quad
H_u=\frac{1}{\sqrt 2}\left(\begin{array}{c} 0\\ \varphi\end{array}\right),
\eeq
with $k$ fixed.
Disregarding $Q$, $u^c$, $d^c$, $e^c$, $H_d$ that play no r\^{o}le during inflation,
the superpotential (\ref{eqn:W}) reads
\beq
W_{\rm inf}
%&=&y_D^{mj}N_m^c L_j H_u+\half M_m N_m^c N_m^c\nn\\
=\half y_D^{mk}N_m^c\varphi^2+\half M_m N_m^c N_m^c.
\label{eqn:Winf}
\eeq
For the K\"{a}hler potential $K\equiv -3\Phi$ we choose a slightly noncanonical form,
\bea
\Phi&=&1-\third\left(|L_k|^2+|H_u|^2+|N_1^c|^2+|N_2^c|^2\right)\nn\\
&&\quad+\frac{\gamma}{2} \left(L_k H_u +c.c.\right)
+\frac{\zeta}{3} \left(|N_1^c|^4+|N_2^c|^4\right)\nn\\
&=&1-\third\left(|\varphi|^2+|N_1^c|^2+|N_2^c|^2\right)\nn\\
&&\quad+\frac{\gamma}{4}\left(\varphi^2+\overline{\varphi}^2\right)
+\frac{\zeta}{3} \left(|N_1^c|^4+|N_2^c|^4\right).
\label{eqn:kahler}
\eea
The terms proportional to $\gamma$ yield nonminimal coupling between the inflaton and the
background curvature, and the other noncanonical terms proportional to $\zeta$ have been
introduced for controlling the inflaton trajectory.
The supergravity scalar potential (in the Jordan frame) is computed from
(\ref{eqn:Winf}) and (\ref{eqn:kahler}) as (see 
%\cite{Kaku:1978nz,Siegel:1978mj,Cremmer:1982en,Ferrara:1983dh,Kugo:1982mr,Kugo:1982cu,Kugo:1983mv}
\cite{superconformal}),
\bea
&&V_F=
\left| y_D^{nk} N_n^c\varphi\right|^2
+A_1\nn\\
&&\qquad-\frac{\left| y_D^{nk}N_n^c(\varphi^2-\frac{3}{2}\gamma|\varphi|^2)+A_2-3W_{\rm inf}\right|^2}{3
-\frac{3}{4}\gamma(\varphi^2+\overline{\varphi}^2)+\frac{9}{4}\gamma^2|\varphi|^2+A_3},
\label{eqn:Vf}
\eea
where the repeated $n$'s are summed over, $k$ is fixed (no sum), and
\bea
A_1&=&\sum_{m=1,2}\frac{\left|\half y_D^{mk}\varphi^2+M_mN_m^c\right|^2}{1-4\zeta\left|N_m^c\right|^2},\nn\\
A_2&=&\sum_{m=1,2}\frac{
\left(\half y_D^{mk}\varphi^2+M_m N_m^c\right){N_m^c}
\left(1-2\zeta\left| N_m^c\right|^2\right)}
{1-4\zeta\left|N_m^c\right|^2},\nn\\
A_3&=&\sum_{m=1,2}\frac{\zeta\left|N_m^c\right|^4}{1-4\zeta\left|N_m^c\right|^2}.
\eea
The Dirac Yukawa coupling is 
\beq
y_D
=\frac{\sqrt 2}{v\sin\beta} \sqrt{M} R\sqrt{D_\nu}U_{\rm MNS}^\dag,
\label{eqn:yDseesaw}
\eeq
%from (\ref{eqn:CIR}) 
and in (\ref{eqn:Vf}) the fields $\varphi$ and $N_m^c$ are understood to be the scalar components.
In the Einstein frame the scalar potential is
\beq
V_{\rm E}=\frac{V_F}{\Phi^2}.
\label{eqn:VE}
\eeq

%%%%%%%%%%%%%%%%%%%%%%%%%%%%%%%%%%%%%%%%%%%%%%%%
\subsection{The inflaton trajectory}
The model of inflation we consider is a system of three complex scalar fields $\varphi$,
$N_1^c$ and $N_2^c$. 
While the dynamics could consequently be quite involved in general, it turns out that
under mild assumptions the model reduces to that of single-field slow roll inflation.
This is by virtue of the noncanonical terms in the K\"{a}hler potential proportional
to $\zeta$.

\begin{figure}[ht]
\includegraphics[width=80mm]{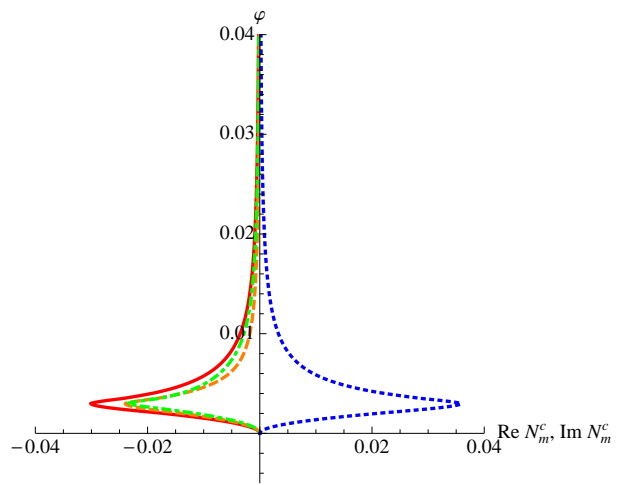}
\caption{The steepest descent trajectory of the scalar potential (\ref{eqn:VE}), for
$M_1=M_2=10^{13}$ GeV, $a=0$, $b=1$, $\delta=0$, $\sigma=0$ and
$k=1$, $\xi=1674$, $\zeta=100$.
The red solid, the orange dashed, the green dot-dashed and the blue dotted curves
respectively shows the values of ${\rm Re} N_1^c$, ${\rm Im} N_1^c$, ${\rm Re} N_2^c$ and 
${\rm Im} N_2^c$.
The slow roll terminates at $\varphi=0.0371$.
The field values are measured in the reduced Planck unit $8\pi G=1$.}
\label{fig:InfTraj}
\end{figure}

\begin{figure}[ht]
\includegraphics[width=80mm]{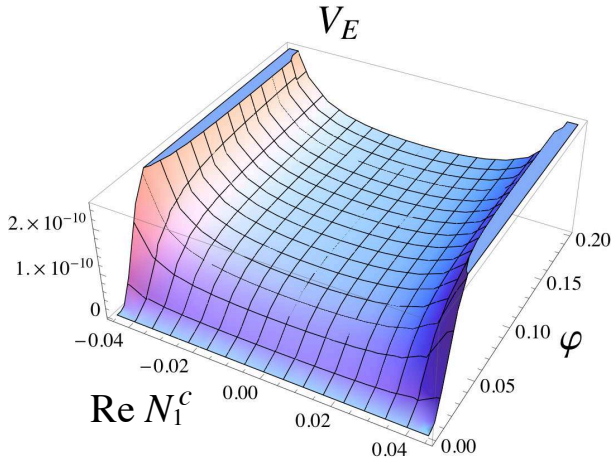}
\caption{The scalar potential $V_{\rm E}$ as a function of $\varphi$ and ${\rm Re} N_1^c$.
The other fields ${\rm Im} N_1^c$, ${\rm Re} N_2^c$, ${\rm Im} N_2^c$ are chosen to take
values along the steepest descent trajectory.}
\label{fig:potential}
\end{figure}

When $M_m$ are large our model is similar to the supersymmetric versions of SM Higgs inflation discussed in \cite{Einhorn:2009bh,Ferrara:2010yw,Ferrara:2010in,Arai:2011nq}.
Let us consider, as an example, the seesaw masses $M_1=M_2=10^{13}$ GeV
and choose for concreteness the normal mass hierarchy of the neutrinos and
$a=0$, $b=1$, $\delta=0$, $\sigma=0$.
Then the Yukawa coupling are found to be
\bea
y_D^{11}=0.0465 + 0.0233 i,\,
&&y_D^{21}=0.0306 - 0.0354 i, \nn\\
y_D^{12}=0.0434 + 0.106 i,\,
&&y_D^{22}=0.139 - 0.0331 i,\nn\\
y_D^{13}=-0.0536 + 0.106 i,\,
&&y_D^{23}=0.139 + 0.0408 i.
\eea
The dynamics of inflation can be studied by examining the steepest descent trajectory of the scalar
potential (\ref{eqn:VE}).
Assuming that inflation takes place in the $L_k$-$H_u$ D-flat direction along the first generation 
lepton supermultiplet ($k=1$) and the e-folding number $N_{e}=60$, we find $\xi=1674$
fixed by the CMB power spectrum (procedure explained in the next subsection).
It is found numerically that the trajectory is along ${\rm Im} \varphi=0$. 
We thus consider only the real values of $\varphi$.
The inflaton trajectory fluctuates in the directions of $N_m^c$ as shown in Fig. \ref{fig:InfTraj},
where the values of the real and imaginary parts of $N_1^c$ and
$N_2^c$ are plotted along the steepest descent trajectory.
The parameter $\zeta$ for the quartic terms in the K\"{a}hler potential is chosen to be a moderate value\footnote{
Being higher order terms in the K\"{a}hler potential, $\zeta$ is expected to be not much larger than 
$\xi$.
For zero or small $\zeta$ the multi-field effects become important.
While the physics of the isocurvature modes and non-Gaussianity arising in the small $\zeta$ case
would also be of interest, we shall not discuss such issues in this paper.}
$\zeta=100$ and the field values are measured in the reduced Planck unit $8\pi G=1$.
While the trajectory seems rather complicated, inflation actually terminates (that is, one of the 
slow roll parameters becomes ${\C O}(1)$) before the complication starts.
In the example considered here the slow roll terminates at $\varphi=0.0371$, 
%$xe=0.0262601$
where $N_1^c$ and $N_2^c$ are still stabilised close to $N_m^c=0$. 
Fig. \ref{fig:potential} shows the shape of the potential as a function of $\varphi$ and 
${\rm Re} N_1^c$, where ${\rm Im} N_1^c$, ${\rm Re} N_2^c$, ${\rm Im} N_2^c$ are taken to be
along the trajectory of Fig. \ref{fig:InfTraj}.
In this case\footnote{
In deriving these values the deviation from $N_1^c$, $N_2^c= 0$ are taken into account but  the dynamics of $N_m^c$ are neglected.
} the primordial tilt and the tensor-to-scalar ratio of the CMB are computed to be
$n_s=0.968$, $r=0.00296$.
The red spectrum with very small $r$ is typical of the nonminimally coupled Higgs inflation.
The horizon exit of the comoving CMB scale takes place at $\varphi=0.318$.
%$xk=0.22486$
In computing the CMB spectrum the r\^{o}le played by $\zeta$ is minor, as $\zeta$ has no effects
once the inflaton trajectory is stabilised along $N_m^c\approx 0$.

%%%%%%%%%%%%%%%%%%%%%%%%%%%%%%%%%%%%%%%%%%
%%%%%%%%%%%%%%%%%%%%%%%%%%%%%%%%%%%%%%%%%%

Setting $N_1^c$, $N_2^c\rightarrow 0$ and restricting $\varphi$ to take real values, the scalar 
potential (\ref{eqn:Vf}) simplifies to
\beq
V_F
=\frac{\lambda}{4} \varphi^4,
%=\frac{(y_D^\dag y_D)_{kk}}{16}\chi^4,
\eeq
where
$\lambda\equiv(y_D^\dag y_D)_{kk}$ is the $k$-$k$ component ($k=1,2,3$) of the 
matrix representation of
\beq
y_D^\dag y_D
=\frac{2}{v^2\sin^2\beta}U_{\rm MNS}\sqrt{D_\nu}^T R^\dag M R\sqrt{D_\nu}U_{\rm MNS}^\dag.
\label{eqn:yDdagyD}
\eeq
The resulting inflationary model is essentially the nonminimally coupled 
$\lambda \phi^4$ model, discussed in \cite{Okada:2010jf}.
We have analysed the above example (corresponding to $\lambda=(y_D^\dag y_D)_{11}=0.00490$)
in this approximation, and found essentially no difference.
For example the predictions for $n_s$ and $r$ are exactly the same within 3 significant digits.
For smaller $M_m$ the effects of nonzero $N_m^c$ become even more negligible, 
as the slow roll terminates at a larger value of
$\varphi$ and the turn around behaviour of $N_m^c$ occurs at the smaller mass scale of $M_m$.

%%%%%%%%%%%%%%%%%%%%%%%%%%%%%%%%%%%%%%%%%%
%%%%%%%%%%%%%%%%%%%%%%%%%%%%%%%%%%%%%%%%%%

%%%%%%%%%%%%%%%%%%%%%%%%%%%%%%%%%%%%%%%%%%%%%%%%
\subsection{Cosmological parameters}
As the single field approximation $N_m^c=0$, $\varphi\in{\B R}$ is sufficiently accurate, 
we shall study the inflationary dynamics within this approximation.
We rescale the inflaton field as
\beq
\chi={\sqrt 2}\varphi,
\eeq
so that the kinetic term is canonically normalised in the Lagrangian
\beq
{\C L}_{\rm J}=\sqrt{-g_{\rm J}}\left[\frac{1+\xi\chi^2}{2} R_{\rm J}
-\half g_{\rm J}^{\mu\nu}\partial_\mu\chi\partial_\nu\chi-V_F\right].
\eeq
Here the subscript J indicates the Jordan frame and
\beq
V_F
%=\frac{(y_D^\dag y_D)_{kk}}{4}\left|\varphi\right|^4
=\frac{\lambda}{16}\chi^4, \quad
%\eeq
%and
%\beq
\xi\equiv\frac{\gamma}{4}-\frac 16.
\eeq
This is the Lagrangian in the Jordan frame where the scalar field is nonminimally coupled to the 
background curvature.
The Lagrangian in the Einstein frame is obtained by the Weyl transformation
$g^{\rm E}_{\mu\nu} =(1+\xi\chi^2) g^{\rm J}_{\mu\nu}$.
The canonically normalised inflaton field $\hat\chi$ in the Einstein frame is related to $\chi$ by
\beq
d\hat\chi=\frac{\sqrt{1+\xi \chi^2+6\xi^2 \chi^2}}{1+\xi \chi^2} d\chi.
\eeq
The potential in the Einstein frame is
\beq
V_{\rm E}=\frac{\lambda}{16}\frac{\chi^4}{(1+\xi \chi^2)^2},
%V_{\rm E}=\frac{(y_D^\dag y_D)_{kk}}{16}\frac{\chi^4}{1+\xi \chi^2}.
\label{eqn:VEeff}
\eeq
in terms of which the slow roll parameters in the Einstein frame are defined as
\beq
\epsilon=\half\left(\frac{1}{V_{\rm E}}\frac{d V_{\rm E}}{d\hat\chi}\right)^2,
\qquad 
\eta=\frac{1}{V_{\rm E}}\frac{d^2 V_{\rm E}}{d\hat\chi^2}.
\eeq

%Cosmological pamameters
The potential $V_{\rm E}$ in the single field approximation (\ref{eqn:VEeff}) is much simpler than 
the original one (\ref{eqn:Vf}) and is determined solely by
$\lambda=(y_D^\dag y_D)_{kk}$ and $\xi$.
%depends on the two real mass parameters $M_1$, $M_2$, the two complex parameters 
%$y_D^{mk}$ (for fixed $k$) as well as on the two K\"{a}hler potential parameters $\gamma$ 
%and $\zeta$, 
For given $\lambda$ and the e-folding number $N_e$, the nonminimal coupling $\xi$ is fixed by 
the CMB power spectrum.
For definiteness we use the maximum likelihood value $\Delta^2_R(k_0)=2.43\times 10^{-9}$
from the 9-year WMAP data \cite{Bennett:2012fp,Hinshaw:2012fq}, where the calibration is set at 
$k_0=0.002 \mbox{ Mpc}^{-1}$.
This is related to the power spectrum ${\C P}_R=V_{\rm E}/24\pi^2\epsilon$ of the curvature perturbation (at the horizon exit of the comoving scale) by 
$\Delta_R^2(k)=\frac{k^3}{2\pi^2}{\C P}_R(k)$.
We denote the value of $\chi$ at the end of the slow roll (characterised by $\max(\epsilon, \eta)= 1$) 
as $\chi=\chi_*$, and the value of $\chi$ at the horizon exit of the comoving CMB scale $k$ as 
$\chi=\chi_k$.
These are related by the e-folding number through 
$N_{e}=\int_{\chi_*}^{\chi_k}d\chi V_{\rm E}({d\hat\chi}/{d\chi})/({d V_{\rm E}}/{d\hat\chi})$.

From the values of the slow roll parameters at the horizon exit of the comoving CMB scale,
the scalar spectral index $n_s\equiv d\ln{\C P}_R/d\ln k=1-6\epsilon+2\eta$ 
and the tensor-to-scalar ratio $r\equiv {\C P}_{\rm gw}/{\C P}_{\rm R}=16\epsilon$ can be computed.
The results\footnote{
These are the results of the tree-level computation.
The renormalisation effects are verified to be negligibly small \cite{Arai:2011nq,Arai:2011aa}.
%as a consequence of the supersymmetry.
}
are summarised in Table \ref{table:numerical}, for $N_e=50$, $60$ and 
several values of $\lambda=(y_D^\dag y_D)_{kk}$.
There exists a lower bound on $\lambda$ set by the minimal coupling limit $\xi\rightarrow 0$.
In this limit the model reduces to chaotic inflation with the quartic potential
$V_{\rm E}=\frac{\lambda}{16}\chi^4$, where $\lambda=1.05\times 10^{-12}$ for $N_e=50$ 
and $\lambda=6.19\times 10^{-13}$ for $N_e=60$, fixed by the value of ${\C P}_R$.
In contrast to the SM
 \cite{CervantesCota:1995tz,Bezrukov:2007ep,Barvinsky:2008ia,DeSimone:2008ei,Bezrukov:2009db,Bezrukov:2010jz,Barvinsky:2009fy,Barvinsky:2009ii,Barbon:2009ya,Burgess:2009ea,Burgess:2010zq,Hertzberg:2010dc,Lerner:2009na,Lerner:2010mq} 
%Barbon:2009ya,Burgess:2010zq,Barvinsky:2009ii,Burgess:2009ea,Lerner:2009na,
%Lerner:2010mq,Hertzberg:2010dc,Bezrukov:2010jz} 
and the supersymmetric 
\cite{Einhorn:2009bh,Ferrara:2010yw,Ferrara:2010in,Arai:2011nq} Higgs inflation models
the nonminimal coupling $\xi$ does not need to be large.
We see from the table that $\xi\lesssim {\C O}(1)$ for $\lambda\lesssim 10^{-9}$.
Hence, at least for these values of $\lambda$ the model is obviously free from the danger of
violating the unitarity bound.
Small $\xi$ is also favoured for avoiding large $R$-parity violation, as discussed later in Section VI.
The Dirac Yukawa coupling $y_D^{mi}$ is related to $a$, $b$, $\delta$, $\sigma$,
$M_1$, $M_2$ through (\ref{eqn:yDseesaw}).
We will discuss the relation between $\lambda$ and these parameters in Section V.
Leptogenesis also constrains certain neutrino mass parameters, which will be discussed in 
Section IV.

%%%%% Table 1 %%%%%

\begin{table}[t]
\begin{center}\begin{tabular}{c|c||ccccc}
$N_e$ & $\lambda\equiv (y_D^\dag y_D)_{kk}$ & $\xi$ & $\chi_*$ & $\chi_k$ & $n_s$ & $r$ \\ \hline
& $0.1$ & 6348 & 0.0135 & 0.106 & 0.962 & 0.00419 \\ 
& $10^{-3}$ & 635 & 0.0426 & 0.335 & 0.962 & 0.00419 \\
& $10^{-5}$ & 63.4 & 0.135 & 1.06 & 0.962 & 0.00420 \\ 
& $10^{-7}$ & 6.26 & 0.424 & 3.33 & 0.962 & 0.00430 \\ 
50 & $10^{-9}$ & 0.555 & 1.28 & 9.94 & 0.961 & 0.00545 \\ 
& $10^{-10}$ & $  0.131 $ & 2.02 & 15.4 & 0.961 & 0.00945 \\ 
& $10^{-11}$ & $ 0.0188 $ & 2.99 & 19.5 & 0.959 & 0.0379 \\
& $5\times 10^{-12}$ & $8.70\times 10^{-3}$ & 3.21 & 20.0 & 0.957 & 0.0691 \\
& $2\times 10^{-12}$ & $2.15\times 10^{-3}$ & 3.39 & 20.2 & 0.951 & 0.165 \\
& $1.05\times 10^{-12}$ & $ 0 $ & 3.46 & 20.3 & 0.942 & 0.311 \\
\hline
& $0.1$ & 7567 & 0.0124 & 0.106 &0.968 & 0.00296 \\ 
& $10^{-3}$ & 757 & 0.0391& 0.335 & 0.968 & 0.00296 \\
& $10^{-5}$ & 75.6 & 0.123 & 1.06 & 0.968 & 0.00297 \\ 
& $10^{-7}$ & 7.48 & 0.389 & 3.33 & 0.968 & 0.00303\\ 
& $10^{-9}$ &  0.676  & 1.18 & 10.0 & 0.968 & 0.00369 \\ 
60 & $10^{-10}$ & 0.168 & 1.90 & 15.9 & 0.968 & 0.00588 \\ 
& $10^{-11}$ & 0.0270 & 2.84 & 20.8 & 0.966 & 0.0203 \\ 
& $5\times 10^{-12}$ & 0.0135 & 3.10 & 21.5 & 0.965 & 0.0356 \\ 
& $2\times 10^{-12}$ & $4.44 \times 10^{-3}$ & 3.32 & 22.0 & 0.962 & 0.0822 \\ 
& $10^{-12}$ & $1.24\times 10^{-3}$ & 3.42 & 22.2 & 0.957 &0.161 \\
& $6.19\times 10^{-13}$ & $0$ & 3.46 & 22.2 & 0.951 & 0.260\end{tabular}
\caption{The nonminimal coupling $\xi$, the inflaton values at the end of the slow roll ($\chi_*$) and at the horizon exit ($\chi_k$), the spectral index $n_s$, and the tensor-to-scalar ratio $r$ for e-folding
$N_e=50$, $60$ and for various different values of $\lambda=(y_D^\dag y_D)_{kk}$.
The coupling $\xi$ is fixed by the amplitude of the curvature perturbation.
The last lines ($N_e=50$, $\lambda=1.05\times 10^{-12}$ and $N_e=60$, $\lambda=6.19\times
10^{-13}$) correspond to minimal coupling.
}\label{table:numerical}
\end{center}
\end{table}

%%%%%%%%%%%%%%%%%%%%%%%%%%%%%%%%%%%%%%%%%%
\begin{figure}[ht]
\includegraphics[width=80mm]{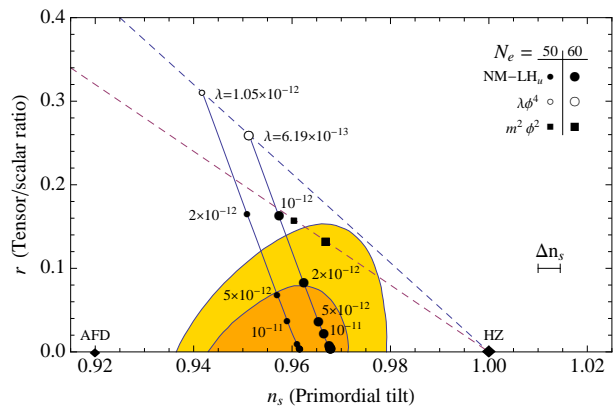}% Here is how to import EPS art
\caption{Plot of the scalar spectral index $n_s$ and the tensor-to-scalar ratio $r$, with the 68\%
and 95\% confidence level contours from the WMAP9$+$eCMB$+$BAO$+H_0$ data 
\cite{Bennett:2012fp,Hinshaw:2012fq}.
Indicated by $\bullet$ are the prediction of our model (NM-$LH_u$) for e-folding numbers 
$N_e=50$ and $60$.
The coupling $\lambda=(y_D^\dag y_D)_{kk}$ is a diagonal element of 
(\ref{eqn:yDdagyD}), which is related to the neutrino mass parameters (see Sec. V).
For comparison, the spectra of the Harrison-Zel'dovich (HZ), the $m^2\phi^2$ minimally coupled 
chaotic inflation model, as well as the A-term-lifed MSSM flat-direction inflation model (AFD)
are also indicated. 
$\Delta n_s$ is the expected Planck accuracy \cite{Planck:2006uk,Ade:2011ah}.
\label{fig:WMAP_Plot}}
\end{figure}
%%%%%%%%%%%%%%%%%%%%%%%%%%%%%%%%%%%%%%%%%%

\subsection{Observational constraints and comparison with other models 
based on supersymmetric seesaw}
We show in Fig.\ref{fig:WMAP_Plot} the predicted values of $n_s$ and $r$ of our model
along with the 68\% and 95\% confidence level contours from the WMAP9$+$eCMB$+$BAO
$+H_0$ data
\cite{Bennett:2012fp,Hinshaw:2012fq}.
The prediction is seen to be consistent with the present data, apart from the regions 
where the Yukawa coupling is extremely small.
The 2-$\sigma$ constraints roughly correspond to $M_m\gtrsim$ TeV, details of which depending
on the other neutrino mass parameters (see Section V).
While the minimally coupled $\lambda \phi^4$ model lies outside the 2-$\sigma$ constraints,
the nonminimally coupled $\lambda \phi^4$, even if the coupling is as small as $\xi\sim10^{-2}$, 
sits well inside the 1-$\sigma$ constraints.
In the figure we also indicate for comparison the Harrison-Zel'dovich spectrum as well as the 
spectra of two other inflationary models arising from the supersymmetric seesaw model. 
These are:
\begin{description}
\item[$\tilde N_R$ chaotic inflation model]
--- Inflation is driven by a right-handed scalar neutrino,
with the seesaw mass scale identified as the inflaton mass
\cite{Murayama:1992ua,Murayama:1993xu,Ellis:2003sq}.
It is essentially the minimally coupled $m^2\phi^2$ chaotic inflation model, but has phenomenological
advantages such as automatic leptogenesis.
The spectrum is marked with {\small $\blacksquare$} in the figure.
\item[A-term inflation model]
--- This model assumes $u^cd^cd^c$, $e^cLL$, or $N_R^c LH_u$ direction of the (singlet-extended) 
MSSM as the inflaton
\cite{Allahverdi:2006iq,Allahverdi:2006we,Allahverdi:2006cx,BuenoSanchez:2006xk}.
The A-term inflation model predicts very small $r$ and $n_s\approx1-4/N_e$.
Since inflation takes place at a low energy scale and the e-folding cannot be large ($N_e\lesssim 50$), we show the case for $N_e=50$ (thus $n_s=0.92$) in the figure, marked with 
$\blacklozenge$ (AFD)
\footnote{
There exists a variant of A-term inflation, called inflection point inflation 
%\cite{Hotchkiss:2011am,Mazumdar:2012qk}.
\cite{inflection}
While this model allows $n_s\geq 0.92$, it has less predictive power on the spectrum.}.
\end{description}
In the figure we also indicate the expected resolution $\Delta n_s\approx 0.0045$ of the Planck 
satellite experiments \cite{Planck:2006uk,Ade:2011ah}.
%Fig.\ref{fig:WMAP_Plot}.
The data from the Planck satellite is soon to be available.
There are more sensitive CMB polarisation experiments planned in the near future 
\cite{Kermish:2012eh,Hazumi:2008zz}, which we expect will put these models to the test
with even higher resolution.
%

%%%%%%%%%%%%%%%%%%%%%%%%%%%%%%%%%%%%%%%%%%%%%%%%
\section{Reheating and baryogenesis}
%%%%%%%%%%%%%%%%%%%%%%%%%%%%%%%%%%%%%%%%%%%%%%%%

%\footnote{
%While larger values of $M_R$ are often favoured in the literature of the seesaw mechanism, 
%its actual bound can be considerably lower \cite{Han:2006ip}.
%}.

% Reheating temperature
The scalar potential (\ref{eqn:VE}) has a minimum at $\varphi=N_m^c=0$
corresponding to the global supersymmetric vacuum.
After the slow roll, the inflaton undergoes coherent oscillations about this minimum and decays into
the SM particles, followed by thermalisation.
The nonminimal coupling can alter the reheating temperature only when $\xi$ is extremely large
and the decay rate is very small \cite{Bassett:1997az,Tsujikawa:1999jh}.
In the model we are discussing the inflaton is in the $L$-$H_u$ direction and the coupling 
to the SM particles is not small.
The nonminimal coupling thus has negligible effects on the reheating process.
The upper bound of the reheating temperature can be estimated as $T_{\rm RH}\lesssim 10^7$ GeV,
assuming the Higgs component decay channel $\varphi\rightarrow b\bar b$ in the conservative
perturbative decay scenario.
The nonperturbative reheating scenario \cite{Allahverdi:2011aj} (with parametric resonance) 
and/or the (s)lepton component decay may lead to a slightly higher upper bound.
Taking into account the effects of the redshift during the time needed for thermalisation, we estimate
the reheating temperature of this scenario to be $T_{\rm RH}\approx10^5 - 10^7$ GeV. 
Below in this section we discuss implication of this reheating temperature on the generation of the 
baryon asymmetry.
The constraints from the R-parity violation will be discussed in Section VI.

% Baryogenesis
The baryon asymmetry (\ref{eqn:YB}) can be accounted for by different mechanisms,
depending on the seesaw mass scale $M_m$.
If the mass scale is smaller than the reheating temperature
$M_m\lesssim T_{\rm RH}$, the right-handed (s)neutrinos thermalise and
their decay generates lepton number asymmetry.
The lepton number is later converted into the baryon number through the 
sphaleron process.
This mechanism is known as thermal leptogenesis \cite{Fukugita:1986hr}.
If the seesaw mass scale is larger than the reheating temperature $M_m\gtrsim T_{\rm RH}$,
on the other hand, the lepton number can be generated by the decay of oscillating sneutrinos
\cite{Murayama:1992ua,Murayama:1993xu,Murayama:1993em}.
We call this mechanism nonthermal leptogenesis.
The generated lepton number is converted into the baryon number, similarly to the
thermal leptogenesis case. 
In addition, as our model includes the MSSM components, the Affleck-Dine mechanism 
\cite{Affleck:1984fy} can be operative.
We shall discuss the thermal and the nonthermal leptogenesis in our model below.

%%%%%%%%%%%%%%%%%%%%%%%%%%%%%%%%%%%%%%%%%%%%%%%%
\subsection{Thermal leptogenesis}
\label{sec:lepto}

In the leptogenesis scenario the out-of-equilibrium decay of the right-handed (s)neutrinos generates
the lepton asymmetry, which is later converted into the baryon asymmetry by the $(B+L)$-violating
sphaleron transitions. 
In the supersymmetric theory the conversion rate is computed to be
\beq
Y_B=-\frac{8N_f+4N_H}{22 N_f+13 N_H} Y_L=-\frac{8}{23}Y_L,
\label{eqn:sphaleron}
\eeq
where
$Y_L$ is the yield (the ratio of the number density to the entropy density) of the leptons, and 
$N_f=3$, $N_H=2$ are the number of the fermion families and the number of the Higgs doublets.

In generic scenarios of leptogenesis, generation of sufficient baryon asymmetry requires
the reheating temperature to be higher than $10^9$ GeV.
This is much higher than the reheating temperature of our scenario, and in 
supersymmetric models the gravitino problem can also be serious.
It has been pointed out, however, that if the two right-handed neutrino masses are nearly degenerate
the CP-asymmetry parameter is enhanced by resonance effects, making the leptogenesis viable
even at lower reheating temperature \cite{Pilaftsis:1997jf,Pilaftsis:2003gt,Flanz:1996fb}.
In the following we assume the resonant leptogenesis scenario with nearly degenerate 
right-handed neutrino masses $M_1\approx M_2$.
Thermalisation of the right-handed neutrinos after reheating also requires 
$M_1$, $M_2\lesssim T_{\rm RH}$.

%%%%%%%%%%%%%%%%%%%%%%%%%%%%%%%%%%%%%%%%%%
\begin{figure}[ht]
\includegraphics[width=80mm]{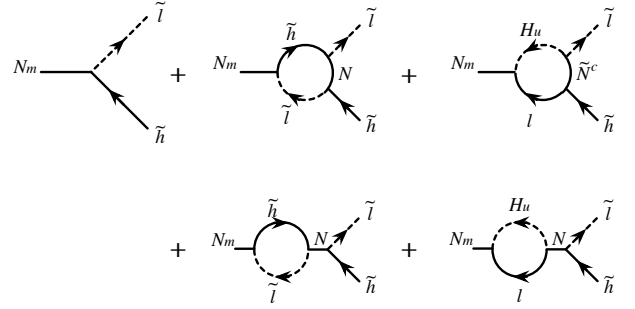}% Here is how to import EPS art
\caption{\label{fig:NDecay}
The diagrams contributing to the decay mode $N_m\rightarrow\tilde\ell+\overline{\widetilde h}$.
}
\end{figure}
%%%%%%%%%%%%%%%%%%%%%%%%%%%%%%%%%%%%%%%%%%

The decay modes of the right-handed (s)neutrinos are
\bea
N_m&\rightarrow&
\widetilde\ell_j+\overline{\widetilde h},\nn\\%\quad
%\widetilde\ell^\dag+\widetilde h,\quad
N_m&\rightarrow&\ell_j+H_u,\nn\\%\quad
%\overline\ell+H_u^\dag,\nn\\
\widetilde{N_m^c}&\rightarrow&
\widetilde\ell_j+H_u,\nn\\%\quad
%\overline\ell+\widetilde h,\nn\\
\widetilde{N_m^c}^\dag&\rightarrow&
\ell_j+\overline{\widetilde h},%\quad
%\widetilde\ell^\dag+H_u^\dag.
\eea
and the CP-asymmetry parameters associated with these processes are defined as
\bea
\varepsilon_m
&\equiv&\frac{
\sum_j\left[\Gamma (N_m\rightarrow\widetilde\ell_j+\overline{\widetilde h})
-\Gamma(N_m\rightarrow\widetilde{\ell}_j^\dag+\widetilde h)\right]}{
\sum_j\left[\Gamma (N_m\rightarrow\widetilde\ell_j+\overline{\widetilde h})
+\Gamma(N_m\rightarrow\widetilde{\ell}_j^\dag+\widetilde h)\right]}\nn\\
&=&\frac{
\sum_j\left[\Gamma (N_m\rightarrow\ell_j+H_u)
-\Gamma(N_m\rightarrow\widetilde{\ell}_j+H_u^\dag)\right]}{
\sum_j\left[\Gamma (N_m\rightarrow\ell_j+H_u)
+\Gamma(N_m\rightarrow\widetilde{\ell}_j+H_u^\dag)\right]}\nn\\
&=&\frac{
\sum_j\left[\Gamma (\widetilde{N_m^c}\rightarrow\widetilde\ell_j+H_u)
-\Gamma(\widetilde{N_m^c}^\dag\rightarrow\widetilde{\ell}_j^\dag+H_u^\dag)\right]}{
\sum_j\left[\Gamma (\widetilde{N_m^c}\rightarrow\widetilde\ell_j+H_u)
+\Gamma(\widetilde{N_m^c}^\dag\rightarrow\widetilde{\ell}_j^\dag+H_u^\dag)\right]}\nn\\
&=&\frac{
\sum_j\left[\Gamma (\widetilde{N_m^c}^\dag\rightarrow\ell_j+\overline{\widetilde h})
-\Gamma(\widetilde{N_m^c}\rightarrow\overline{\ell}_j+\widetilde h)\right]}{
\sum_j\left[\Gamma (\widetilde{N_m^c}^\dag\rightarrow\ell_j+\overline{\widetilde h})
+\Gamma(\widetilde{N_m^c}\rightarrow\overline{\ell}_j+\widetilde h)\right]}.\nn\\
\eea
Here, $\widetilde{N_m^c}$ denotes the right-handed scalar neutrinos, $\widetilde h$ denotes
the (up-type) higgsinos, and $\ell$, $\widetilde\ell$ are the components of the lepton and scalar 
lepton doublets.
We follow the notation of \cite{Plumacher:1997ru}.
The CP asymmetry parameters are computed from the interference of the tree and one-loop 
diagrams.
In Fig. \ref{fig:NDecay} we show the diagrams contributing to one of the decay modes
$N_m\rightarrow\tilde\ell_j+\overline{\widetilde h}$.
The resulting formula for the CP-asymmetry parameters is
\beq
\varepsilon_m\equiv
\sum_{n\neq m}
\frac{-{\rm Im}[(y_Dy_D^\dag)^2_{mn}]}{(y_Dy_D^\dag)_{mm}(y_Dy_D^\dag)_{nn}}
\frac{M_m \Gamma_n}{M_n^2}\left(\half V_n+S_n\right),
\label{eqn:epsilon}
\eeq
where $\Delta M_{mn}^2=M_m^2-M_n^2$, and
\bea
V_n&=&\frac{M_n^2}{M_m^2}\ln\left(1+\frac{M_m^2}{M_n^2}\right),\\
S_n&=&\frac{M_n^2\Delta M_{nm}^2}{(\Delta M_{nm}^2)^2+M_m^2\Gamma_n^2},
\eea
corresponding respectively to the vertex corrections (such as the 2nd and 3rd diagrams of 
Fig. \ref{fig:NDecay}) and the self-energy corrections (the 4th and 5th diagrams). 
The decay width is written as 
\beq
\Gamma_n=\frac{(y_Dy_D^\dag)_{nn}}{4\pi}{M_n}.
\label{eqn:Gamma_N}
\eeq

\begin{figure}[ht]
\includegraphics[width=42mm]{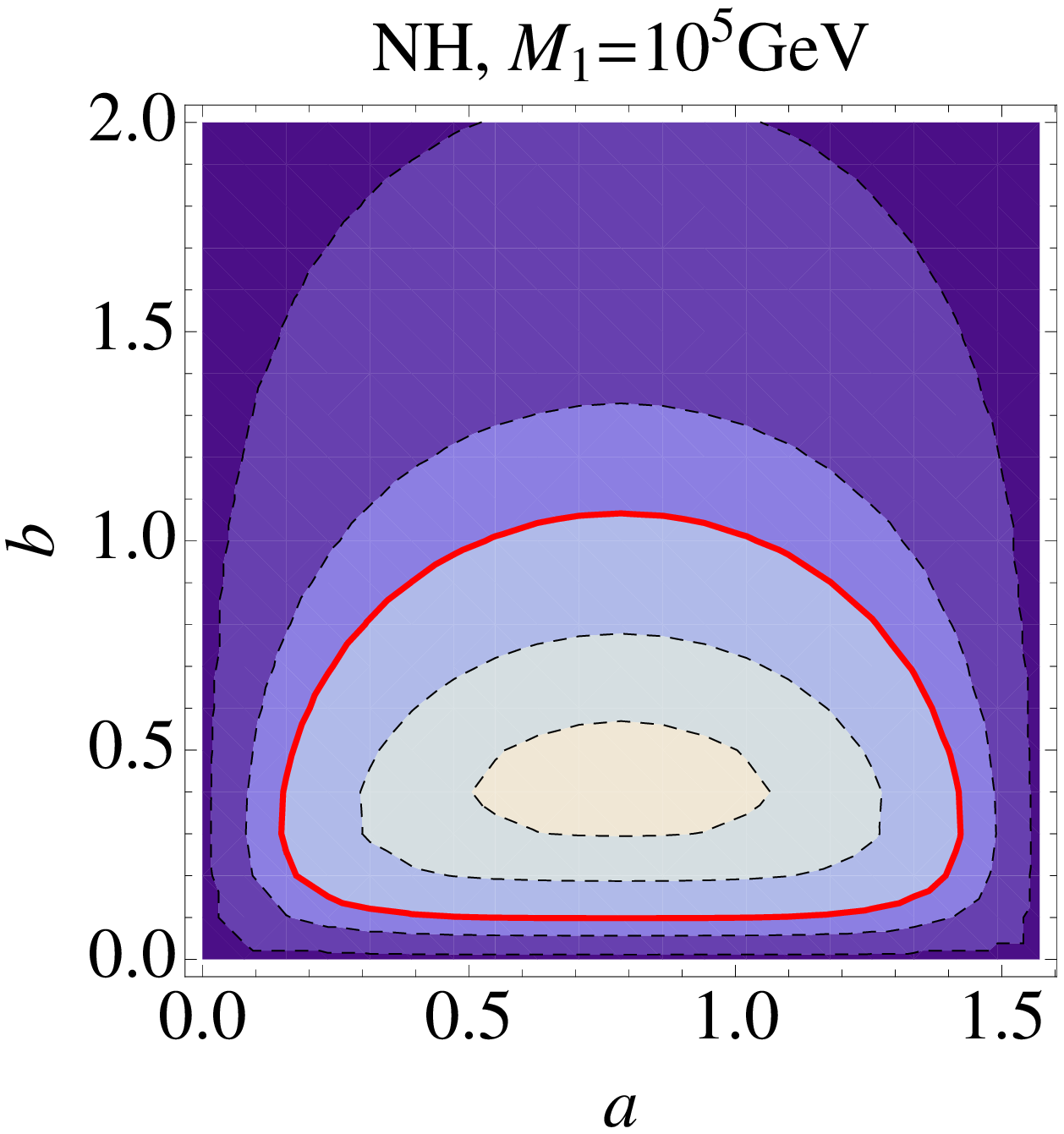}% Here is how to import EPS art
\includegraphics[width=42mm]{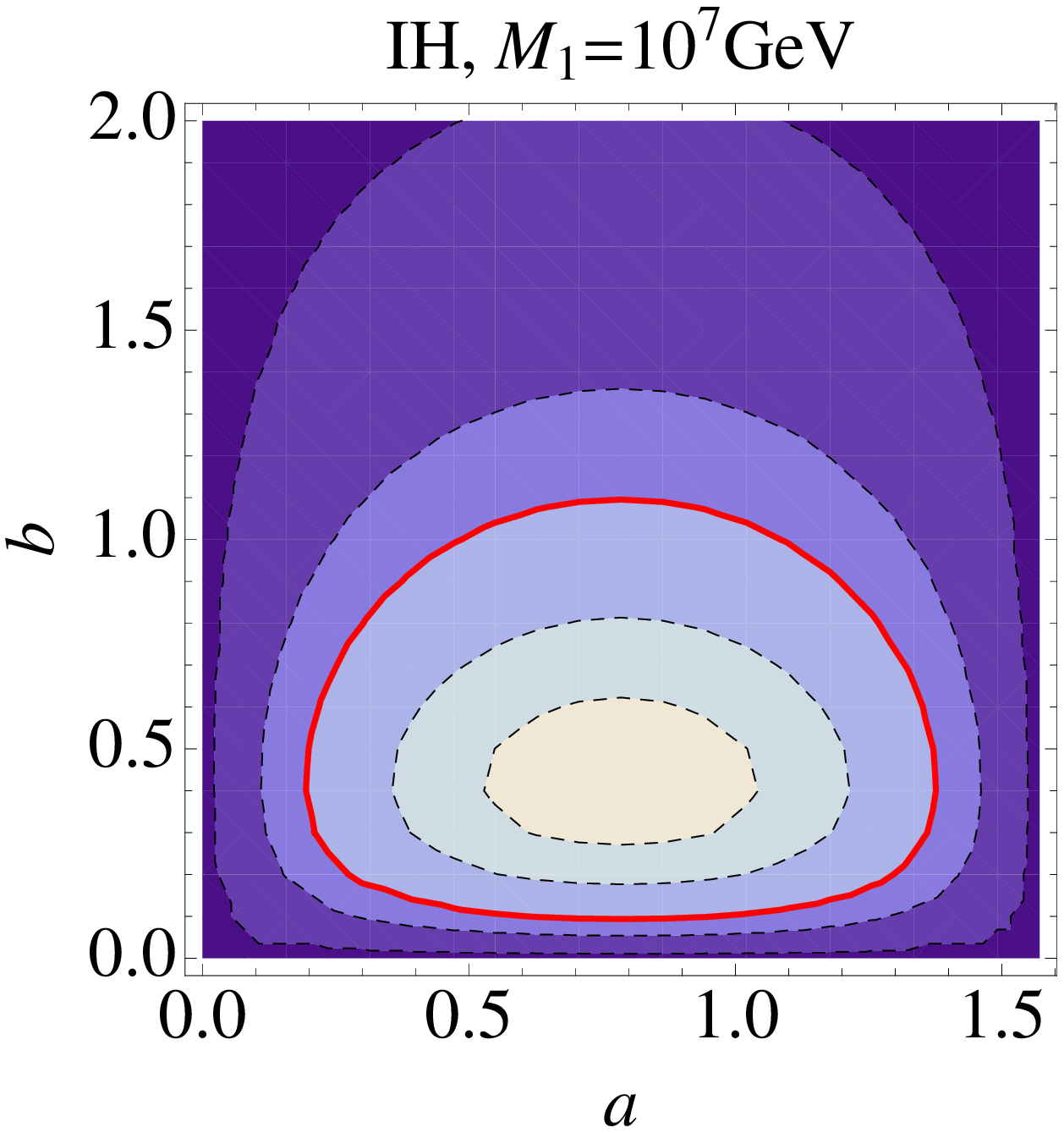}
\caption{\label{fig:BAU}
The neutrino mass parameters $a$ and $b$ yielding the observed value of the baryon
asymmetry $Y_B=8.7\times 10^{-11}$ (the red contour curves).
The left panel shows the case $M_1=10^5$ GeV with the normal mass hierarchy, and 
the right panel is for $M_1=10^7$ GeV with the inverted mass hierarchy.
In both panels the contours are
$Y_B=2\times 10^{-10}$,
$1.5\times 10^{-10}$,
$8.7\times 10^{-11}$, 
$5\times 10^{-11}$,
$1\times 10^{-11}$
from inside.
We have chosen $(M_2-M_1)/M_1=10^{-7}$.
}
\end{figure}

The lepton asymmetry $Y_L$, and hence the baryon asymmetry through the conversion 
(\ref{eqn:sphaleron}), are found by solving the Boltzmann equations.
We summarise the Boltzmann equations and related formulae in the Appendix. 
Here we discuss the results and features that are relevant to our model.
We first notice that the CP phases $\delta$ and $\sigma$ do not affect the baryon asymmetry. 
This is due to the fact that the MNS matrices cancel in the product of the Yukawa couplings appearing in (\ref{eqn:epsilon}),
\beq
y_D y_D^\dag=\frac{2}{v^2\sin^2\beta}\sqrt{M} R \widetilde{D}_{\nu}R^\dag\sqrt{M}.
\label{eqn:yydag}
\eeq
Here,
$\widetilde{D}_{\nu}={\rm diag} (m_2, m_3)$ for the normal mass hierarchy and
$\widetilde{D}_{\nu}={\rm diag} (m_1, m_2)$ for the inverted mass hierarchy. 
The baryon number, on the other hand, depends on the parameters $a$, $b$ in 
(\ref{eqn:R}), as well as on the right-handed neutrino masses $M_1$, $M_2$. 
In Fig.\ref{fig:BAU} we show the yield of the baryon asymmetry $Y_B$, as $a$ and $b$ are
varied.
The left panel shows the result for the normal mass hierarchy with $M_1=10^5$ GeV,
and the right panel is for the inverted mass hierarchy with $M_1=10^7$ GeV.
In both panels the mass difference is chosen to be $M_2-M_1=10^{-7}\times M_1$.
%, and the contours are
%$Y_B=2\times 10^{-10}$,
%$1.5\times 10^{-10}$,
%$8.7\times 10^{-11}$, 
%$5\times 10^{-11}$,
%$1\times 10^{-11}$
%from inside.
The red contour curves indicate $Y_B=8.7\times 10^{-11}$ corresponding to the observed value.

The dependance of the baryon asymmetry of the universe on these parameters can be understood
as follows. 
The baryon asymmetry generated through thermal leptogenesis is proportional to
the CP violation parameter $\varepsilon_m$ as \cite{Buchmuller:2002rq,Buchmuller:2004nz}
\beq
Y_B\approx \kappa\frac{\varepsilon_m}{g_*},
\eeq
where $g_*\approx 200$ is the number of degrees of freedom during leptogenesis and
$\kappa\lesssim 1$ is the efficiency factor depending on details of the Boltzmann equations.
In the resonant leptogenesis with nearly degenerate right-handed neutrino masses
$M_1\approx M_2$ we find $V_n\approx \ln 2\ll S_n$.
The maximal enhancement of the CP asymmetry parameter $\varepsilon_m$ occurs
when the decay width of either of the right-handed (s)neutrinos becomes close to the mass 
difference\footnote{
While this might seem enormous fine tuning, it can happen naturally as a result of renormalisation 
group effects \cite{GonzalezFelipe:2003fi}. See also \cite{Okada:2012fs}.},
$\Delta M_{nm}^2\approx M_m\Gamma_n$.
Parametrizing the mass difference as
$\Delta M_{21}^2=\alpha M_1\Gamma_2$,
the CP asymmetry parameters read
\bea
\varepsilon_1&\approx&
-\frac{\alpha}{1+\alpha^2}
\frac{{\rm Im}\left[(y_Dy_D^\dag)_{12}^2\right]}{(y_Dy_D^\dag)_{11}
(y_Dy_D^\dag)_{22}},\nn\\
\varepsilon_2&\approx& 
\frac{(\alpha^2+1)\Gamma_1\Gamma_2}{\Gamma_1^2+\alpha^2\Gamma_2^2}
\varepsilon_1.
\eea
For the (nearly) degenerate right-handed neutrino masses, the matrix $\sqrt M$ in (\ref{eqn:yydag})
becomes proportional to the identity.
Then 
${\rm Im}\left[(y_Dy_D^\dag)_{12}^2\right]/(y_Dy_D^\dag)_{11}
(y_Dy_D^\dag)_{22}$ is written as
\beq
\frac{2(m_2^2-m_3^2) \sin 2a \sinh 2b}{(m_2-m_3)^2\cos^2 2a-(m_2+m_3)^2\cosh^2 2b}
\label{eqn:ImyydNH}
\eeq
for the normal mass hierarchy and 
\beq   
\frac{2(m_1^2-m_2^2) \sin 2a \sinh 2b}{(m_1-m_2)^2\cos^2 2a-(m_1+m_2)^2\cosh^2 2b}
\label{eqn:ImyydIH}
\eeq
for the inverted mass hierarchy.
These have maxima at $a=\pi/4\approx 0.785$ and $b=\ln (3+2\sqrt 2)/4\approx 0.441$,
with maximal value
$(m_3-m_2)/(m_3+m_2)\approx 0.704$ for (\ref{eqn:ImyydNH}) and 
$(m_2-m_1)/(m_2+m_1)\approx 7.93\times 10^{-3}$ for (\ref{eqn:ImyydIH}). 
Note that $\varepsilon_m$ do not depend on the seesaw mass scale $M_m$ except through 
$\alpha$, $\Gamma_1$ and $\Gamma_2$.
By adjusting the parameters $a$, $b$ and $\alpha$ that are not constrained by observation,
it is always possible to reproduce the baryon asymmetry $Y_B=8.7\times 10^{-11}$.
% for the range $M_1$, $M_2$ =...

%%%%%%%%%%%%%%%%%%%%%%%%%%%%%%%%%%%%%%%%%%
%%%%%%%%%%%%%%%%%%%%%%%%%%%%%%%%%%%%%%%%%%

%%%%%%%%%%%%%%%%%%%%%%%%%%%%%%%%%%%%%%%%%%%%%%%%
\subsection{Nonthermal leptogenesis due to decaying right-handed scalar neutrinos}

The mechanism described above operates when the right-handed neutrino masses $M_1$, $M_2$ 
are smaller than the reheating temperature which we assume in our scenario to be in the 
range $T_{\rm RH}=10^{5}-10^{7}$ GeV.
For larger $M_1$, $M_2$ the thermal leptogenesis scenario is not applicable as the right-handed 
(s)neutrinos do not thermalise.
It is nevertheless possible to generate sufficient baryon asymmetry due to the 
decay of the right-handed sneutrinos that have acquired expectation values during inflation.
%\cite{Murayama:1992ua,Murayama:1993xu,Ellis:2003sq,Murayama:1993em}.
In this sense our model is somewhat similar to the inflation model driven by the right-handed
scalar neutrinos \cite{Murayama:1992ua,Murayama:1993xu,Ellis:2003sq}.
Indeed, for large seesaw scales the right-handed scalar neutrinos have noticeable contribution
to the dynamics after the slow-roll, as shown in Fig.\ref{fig:InfTraj} for $M_1=M_2=10^{13}$ GeV.
Note, however, that the right-handed scalar neutrinos do not have to be involved in the inflationary 
dynamics.

In the supersymmetric SM including the right-handed neutrinos leptogenesis is automatic
as long as the Hubble scale during inflation $H_{\rm inf}$ is larger than the right-handed neutrino
mass scale \cite{Murayama:1993em}, which is always the case in our model.
%This condition is always satisfied in our model.
Take, for example, the case of $M_1=M_2=10^{13}$ GeV and $N_e=60$ discussed in 
Section III B, which gives $r=0.00296$. 
Using the CMB power spectrum $\Delta_R^2(k_0)=2.43\times 10^{-9}$, the Hubble parameter
during inflation can be expressed in terms of the tensor-to-scalar ratio $r=16\epsilon$ as
$H_{\rm inf}=2.67\times\sqrt r\times 10^{14}$ GeV.
With $r=0.00296$ the Hubble parameter during inflation is larger than the right-handed neutrino
masses,
$H_{\rm inf}\approx 1.45\times 10^{13}$ GeV $> M_1$, $M_2$.
The seesaw relation (\ref{eqn:seesaw}) with the Yukawa coupling $\lesssim {\C O}(1)$ constrains
$M_1$, $M_2\lesssim 10^{13}$ GeV, and 
for smaller $M_1$, $M_2$ the model yields larger $r$ and hence larger $H_{\rm inf}$, so the
inequality $H_{\rm inf}> M_1$, $M_2$ is satisfied more safely.

In this nonthermal leptogenesis scenario the right-handed scalar neutrinos acquire expectation
values $\langle\widetilde N\rangle$, either classically by the nontrivial inflaton trajectory, or 
due to the quantum fluctuation during inflation.
This $\langle\widetilde N\rangle$ can be regarded as the initial value of the right-handed scalar 
neutrinos that oscillate and decay.
We shall assume $\langle\widetilde N\rangle\lesssim M_{\rm Pl}$, where
$M_{\rm Pl}$ is the Planck mass.
Along with the decay of the right-handed scalar neutrinos the lepton asymmetry is generated, 
which is later converted into the baryon asymmetry by the $B-L$ conserving sphaleron process.
The process of the decay actually depends on the decay rate of the inflaton $\Gamma_\varphi$
and that of the right-handed neutrinos $\Gamma_n$.
%Assuming $\langle\widetilde N\rangle\lesssim M_{\rm Pl}$,
When (i) the scalar neutrino decay rate is larger than the inflaton decay rate 
$\Gamma_n>\Gamma_\varphi$, the scalar neutrinos decay during the reheating.
When (ii) $\Gamma_n<\Gamma_\varphi<\Gamma_n (M_{\rm Pl}/\langle \widetilde N\rangle)^4$, 
the scalar neutrinos decay after the reheating but do not dominate the energy density of the universe. 
When (iii) the scalar neutrino decay rate is much smaller 
$\Gamma_\varphi>\Gamma_n(M_{\rm Pl}/\langle\widetilde N\rangle)^4$, 
the right-handed scalar neutrinos dominate the universe before they decay. 
In our model, the scalar neutrino decay rate\footnote{
Parametric resonance may enhance the decay rate.
We do not consider such effects here.} is (\ref{eqn:Gamma_N}) and
the inflaton decay rate is estimated as $\Gamma_\varphi\approx 4$ MeV, assuming the Higgs 
$\rightarrow b\bar b$ decay channel and the Higgs mass $\approx$ 125 GeV.
It can be shown that in our scenario the case (iii) never occurs, as long as the inflaton trajectory
is controlled to be nearly straight by the noncanonical K\"{a}hler terms. 
The threshold between (i) and (ii) is around $M_1\approx M_2\approx 10^{6.5}$ GeV, 
for typical values of the mass parameters $a=b=1$ and for the normal or inverted mass hierarchy.

In both (i) and (ii) cases the yield of the lepton asymmetry is estimated as
\cite{Murayama:1993em}
\beq
Y_L=\varepsilon_m\frac{\langle\widetilde N\rangle^2}{M_m M_{\rm Pl}}
\left(\frac{\Gamma_\varphi}{M_{\rm Pl}}\right)^\half,
\eeq
and the baryon asymmetry follows from (\ref{eqn:sphaleron}). 
Assuming
\beq
\langle\widetilde N\rangle^2\simeq \frac{3 H_{\rm inf}^4}{8\pi^2 M_m^2}
\eeq 
generated by the quantum fluctuations\footnote{
%For large right-handed neutrino masses $M_1$, $M_2\sim10^{13}$ GeV 
If $\langle\widetilde N\rangle$ is set by the nontrivial inflaton trajectory as in the case of
Fig.\ref{fig:InfTraj}, the value of $\langle\widetilde N\rangle$ depends on $\zeta$.
Then the lepton number generated through coherent oscillation and decay of the scalar neutrinos
also depends on the parameter $\zeta$.}, 
the observed baryon asymmetry (\ref{eqn:YB}) with the condition for the CP asymmetry
parameter $|\varepsilon_m|\lesssim 1$ leads to a mild upper bound on the seesaw mass scale
$M_m\lesssim 10^{12}$ GeV.

%%%%%%%%%%%%%%%%%%%%%%%%%%%%%%%%%%%%%%%%%%%%%%%%
\section{The inflaton self coupling and the neutrino mass parameters}
%%%%%%%%%%%%%%%%%%%%%%%%%%%%%%%%%%%%%%%%%%%%%%%%
%
Our model of inflation is well approximated by the nonminimally coupled
single field $\lambda\phi^4$ model.
We have seen in Section III that the inflaton self coupling $\lambda=(y_D^\dag y_D)_{kk}$
is related to the CMB spectrum, and argued that in the near future the value of $\lambda$ will
be constrained more severely by the observation. 
As $\lambda$ is constructed from components of the Dirac Yukawa coupling, it is
determined by the neutrino mass parameters. 
In this section we describe the relation between $\lambda$ and these parameters.

We shall assume nearly degenerate right-handed neutrino mass parameters $M_1\approx M_2$,
which is favoured for successful thermal leptogenesis as we discussed in the previous section.
The matrix $M={\rm diag}(M_1, M_2)$ is then proportional to the identity matrix and the 
relation (\ref{eqn:yDdagyD}) becomes
\beq
y_D^\dag y_D
\approx
\frac{2 M_1}{v^2\sin^2\beta}U_{\rm MNS}\sqrt{D_\nu}^T R^\dag R\sqrt{D_\nu}U_{\rm MNS}^\dag.
\label{eqn:ydy_2}
\eeq
Thus $(y_D^\dag y_D)_{kk}$ is proportional to the seesaw mass parameter $M_1$.
Essentially, the inflaton self coupling is determined by the seesaw mass scale
\cite{Arai:2011aa}.
Further complexity arises from the dependence on the other parameters $b$, %the Dirac phase 
$\delta$ and %the Majorana phase 
$\sigma$, which are not constrained by the present observation.

\begin{figure}[ht]\begin{center}
%\begin{eqnarray*}
%\begin{array}{cccc}
\includegraphics[width=42mm]{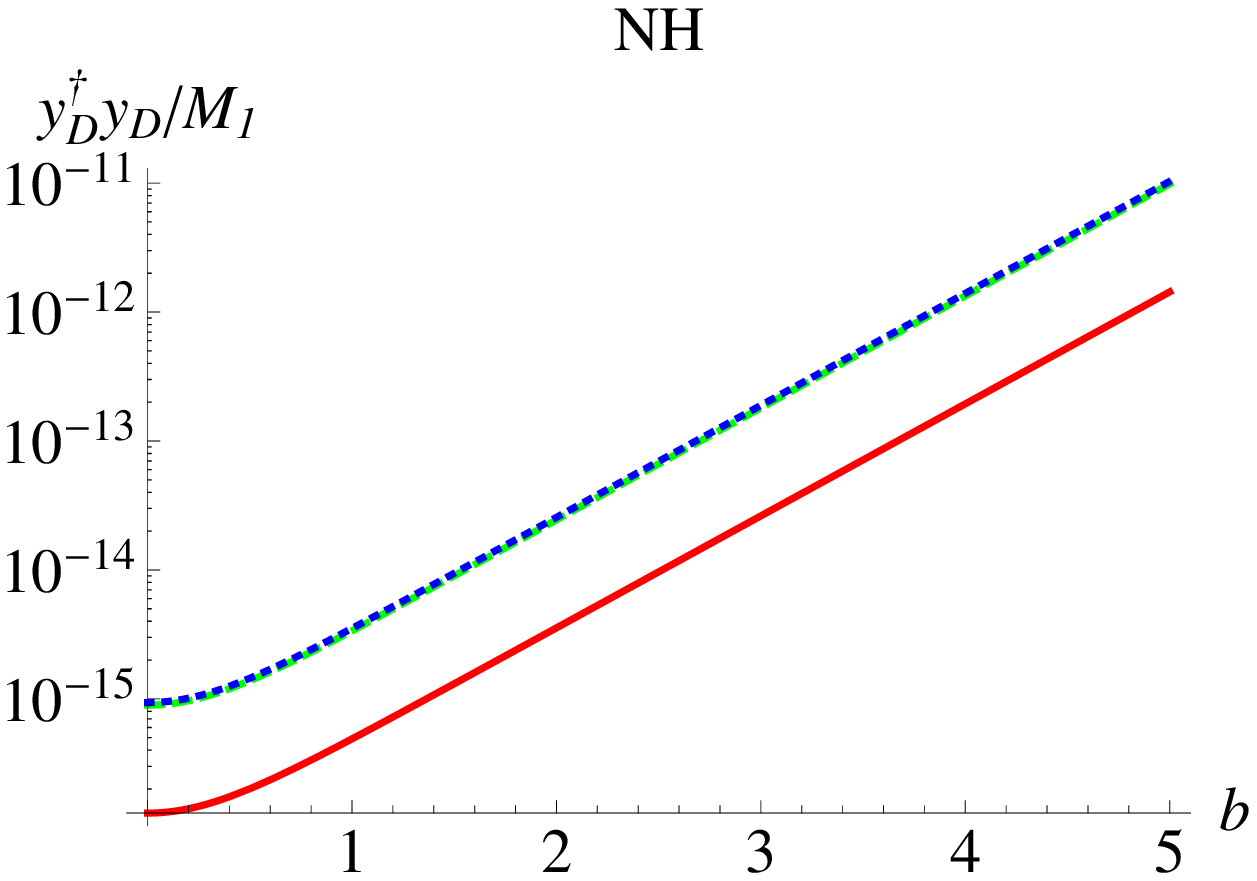} % Here is how to import EPS art
\includegraphics[width=42mm]{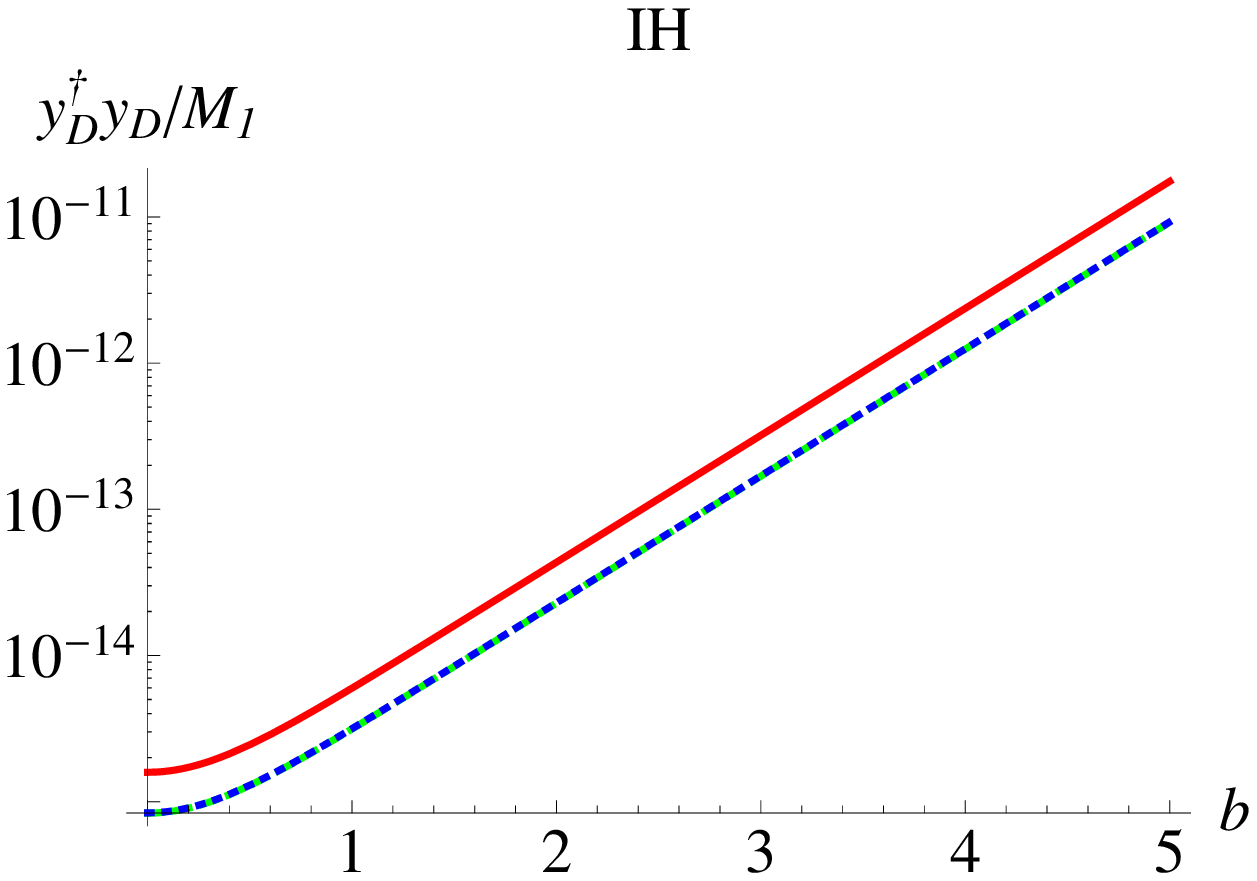}
%\end{array}
%\end{eqnarray*}
\caption{\label{fig:yDonb}
Plot of $\lambda/M_1\equiv (y_D^\dag y_D)_{kk}/M_1$ (in GeV${}^{-1}$) as functions of $b$.
The Dirac and Majorana phases are chosen to be zero.
The left panel is for the normal mass hierarchy and the right is for the inverted mass hierarchy.
On each panel, the red solid, the green dashed, and the blue dotted curves indicate
%$\lambda\equiv (y_D^\dag y_D)_{11}$,
%$\lambda\equiv (y_D^\dag y_D)_{22}$,
%$\lambda\equiv (y_D^\dag y_D)_{33}$
$k=1,2,3$, respectively.
The $k=2$ and $k=3$ curves overlie each other.
}
\end{center}\end{figure}

%\subsection{Dependence on $a$ and $b$}
{\em Dependence on $a$ and $b$}. --- 
We first notice that $\lambda$ does not depend on the parameter $a$.
This is easily seen, as the product $R^\dag R$ appearing in (\ref{eqn:ydy_2}) is written as
\beq
R^\dag R=
\left(
\begin{array}{cc}
 \cosh 2b & i \sinh 2b \\
 -i \sinh 2b & \cosh 2b \\
\end{array}
\right).
\eeq
The coupling $\lambda$ monotonically increases with $b>0$.
We show the behaviour of $(y_D^\dag y_D)_{kk}/M_1$ as functions of $b$ in Fig. \ref{fig:yDonb}, 
for $k=1,2,3$, with the CP violating phases set to be $\delta=0$, $\sigma=0$.
The self coupling $\lambda$ changes by an order as $b$ is shifted by 1;
the dependence on $b$ is thus significant.

%%%%%%%%%%%%%%%%%%%%%%%%%%%%%%%%%%%%%%%%%%

%%%%%%%%%%%%%%%%%%%%%%%%%%%%%%%%%%%%%%%%%%

\begin{figure}[ht]\begin{center}
\includegraphics[width=42mm]{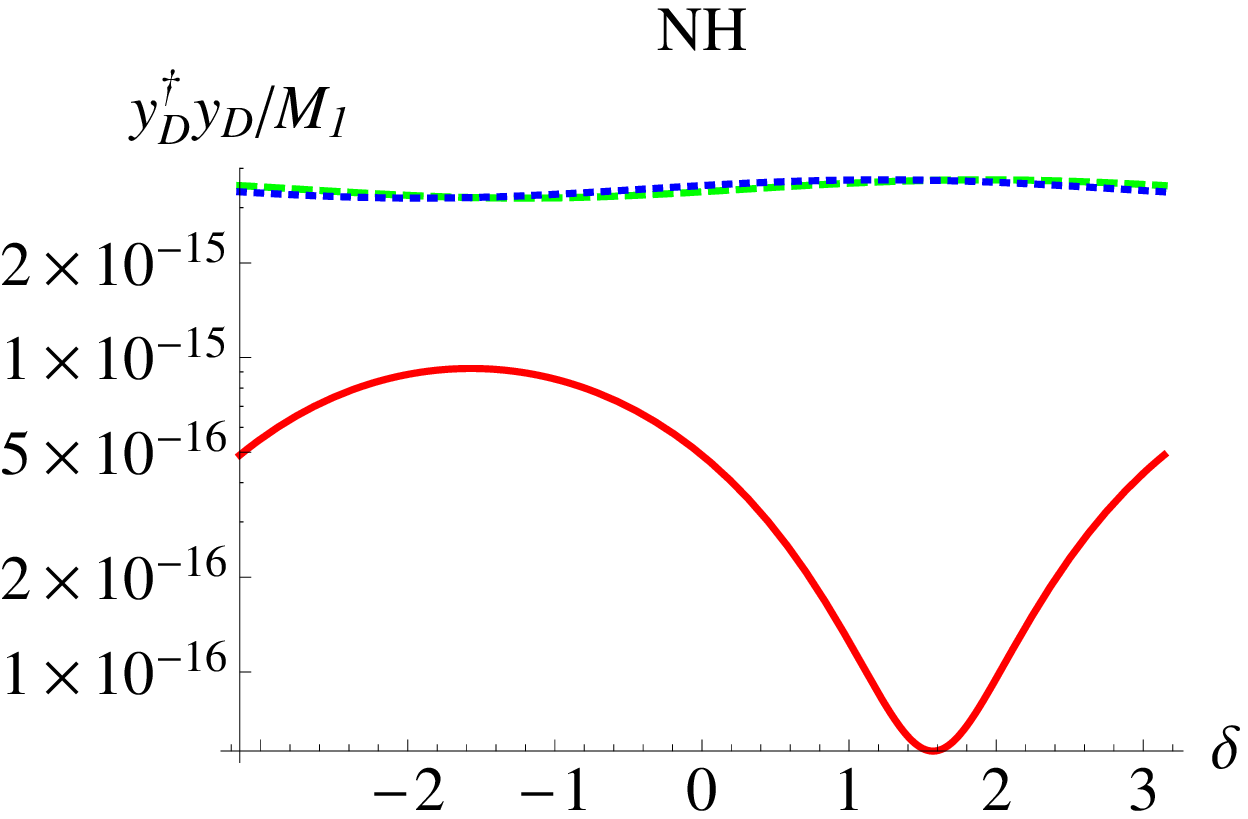}% Here is how to import EPS art
\includegraphics[width=42mm]{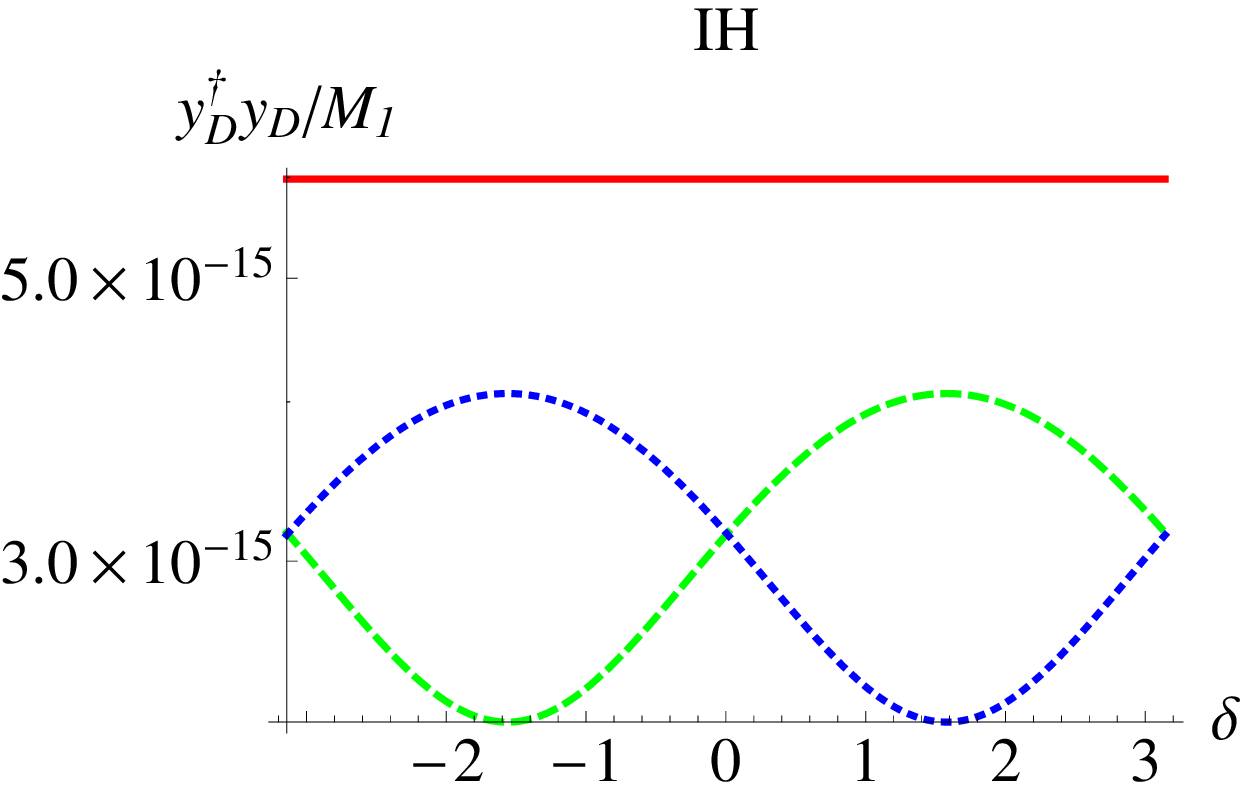}
\caption{
Plot of $\lambda/M_1\equiv (y_D^\dag y_D)_{kk}/M_1$ for $k=1$ (the red solid curves), 
$k=2$ (green dashed) and $k=3$ (blue dotted), as functions of the Dirac phase $\delta$.
The unit of $\lambda/M_1$ is in GeV${}^{-1}$.
The left panel shows the case for the normal mass hierarchy and the right panel is for the inverted 
mass hierarchy.
We have chosen $b=1$ and $\sigma=0$. 
}
\label{fig:yDondelta}
\end{center}
\end{figure}

%\subsection{Dependence on the CP violating Dirac phase $\delta$}
{\em Dependence on the Dirac phase $\delta$}. ---
The dependence of the inflaton self coupling $\lambda$ on the Dirac phase $\delta$ is
shown in Fig. \ref{fig:yDondelta},
where $(y_D^\dag y_D)_{kk}/M_1$ is plotted for $-\pi\leq\delta<\pi$.
We have chosen $b=1$ and $\sigma=0$.
In the case of the normal mass hierarchy the $k=1$ component is more susceptible than $k=2,3$,
whereas for the inverted mass hierarchy $k=2,3$ are more susceptible than $k=1$.

%%%%%%%%%%%%%%%%%%%%%%%%%%%%%%%%%%%%%%%%%%
%%%%%%%%%%%%%%%%%%%%%%%%%%%%%%%%%%%%%%%%%%

\begin{figure}[ht]\begin{center}
\includegraphics[width=42mm]{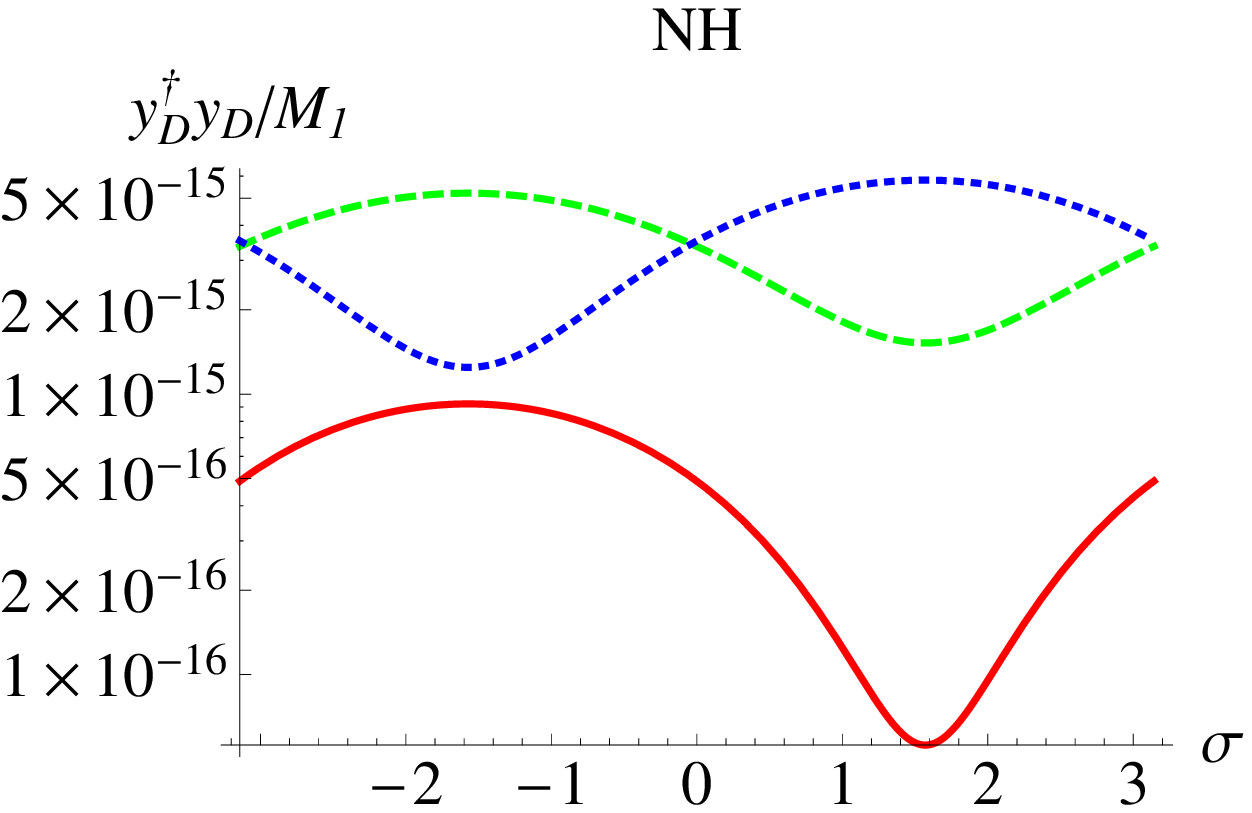}% Here is how to import EPS art
\includegraphics[width=42mm]{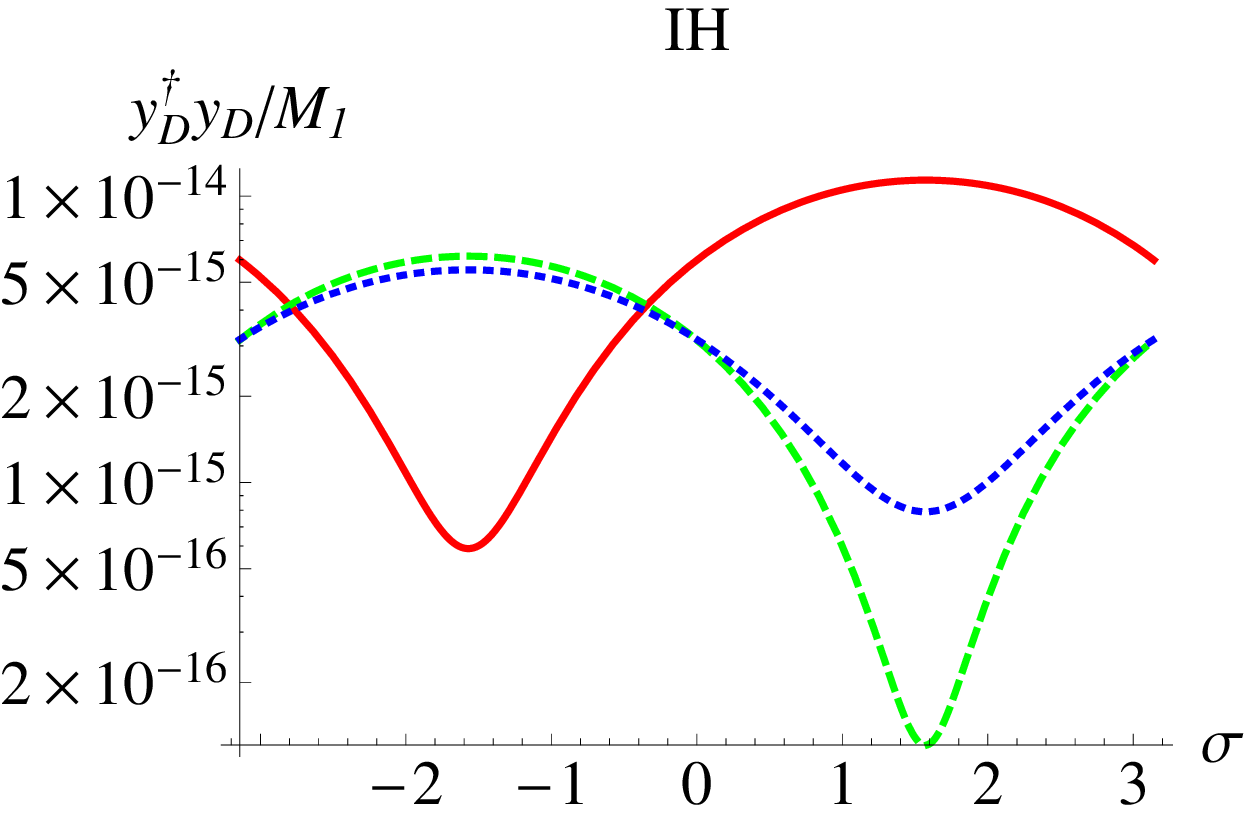}
\caption{
Plot of $\lambda/M_1\equiv (y_D^\dag y_D)_{kk}/M_1$ for $k=1$ (the red solid curves), 
$k=2$ (green dashed), $k=3$ (blue dotted), as functions of the Majorana phase $\sigma$.
The unit of $\lambda/M_1$ is in GeV${}^{-1}$.
The left panel shows the case for the normal mass hierarchy and the right panel is for the inverted 
mass hierarchy.
We have chosen $b=1$ and $\delta=0$. 
}
\label{fig:yDonsigma}
\end{center}
\end{figure}

%\subsection{Dependence on the CP violating Majorana phase $\sigma$}
{\em Dependence on the Majorana phase $\sigma$}. ---
The inflaton self coupling $\lambda$ also depends on the Majorana phase $\sigma$.
In Fig. \ref{fig:yDonsigma} we show the behaviour of $(y_D^\dag y_D)_{kk}/M_1$ as 
$\sigma$ is varied from $-\pi$ to $\pi$.
We have chosen $b=1$ and $\delta=0$.

%%%%%%%%%%%%%%%%%%%%%%%%%%%%%%%%%%%%%%%%%%
%%%%%%%%%%%%%%%%%%%%%%%%%%%%%%%%%%%%%%%%%%

To summarise, the inflaton self coupling $\lambda$ depends on $b$, $\delta$ and $\sigma$
but not on $a$.
These parameters are not constrained by present experiments and thus introduce ambiguity
in prediction of the model.
Consider for instance the 2-$\sigma$ contour in Fig.\ref{fig:WMAP_Plot}. 
For $N_e=60$ this gives a constraint $\lambda=(y_D^\dag y_D)_{kk}\gtrsim 10^{-12}$ which 
corresponds to $M_1\approx M_2\gtrsim 2$ TeV for $b=1$, $\delta=\sigma=0$, $k=1$ in
the normal mass hierarchy (Fig.\ref{fig:yDondelta}).
If we take $\delta=1.6$ instead of $\delta=0$, the $2$-$\sigma$ of CMB gives
$M_1\approx M_2\gtrsim 20$ TeV.

%%%%%%%%%%%%%%%%%%%%%%%%%%%%%%%%%%%%%%%%%%%%%%%%
\section{The effects of R-parity violation}
%%%%%%%%%%%%%%%%%%%%%%%%%%%%%%%%%%%%%%%%%%%%%%%%
%

While the seesaw model superpotential (\ref{eqn:W}) preserves the R-parity
(assuming the odd parity for the right-handed neutrino superfield),
the K\"{a}hler potential (\ref{eqn:kahler}) breaks it by the $\gamma_i L_i H_u$ terms.
Small R-parity violation is known to be harmless, and it can often be beneficial
\cite{Dreiner:1997uz,Barbier:2004ez,Buchmuller:2007ui}.
It can nevertheless lead to difficulties and the consequences of introducing such terms need to
be checked carefully.
In this section we discuss viability of the scenario focusing on the effects of R-parity violation.
We assume the canonical form of the K\"{a}hler terms for all the other chiral multiplets.

An immediate consequence of the $\gamma_i L_i H_u$ terms in the K\"{a}hler potential is that 
the the Lagrangian includes the following terms:
\bea
{\C L}&\supset&\int d^4\theta\phi^\dag\phi\left(
L_i^\dag L_i + H_u^\dag H_u +\gamma_i L_i H_u +h.c.
\right)\nn\\
&=&\int d^4\theta\left(
L_i^\dag L_i + H_u^\dag H_u +\frac{\phi^\dag}{\phi}\gamma_i L_i H_u +h.c.
\right),
\label{eqn:Leff}
\eea
where $\phi$ is the compensator in the superconformal formalism.
To go from the first to the second line we have rescaled
$L_i, H_u\rightarrow \phi^{-1}L_i, \phi^{-1} H_u$.
When the supersymmetry is broken the compensator acquires an expectation value 
$\langle\phi\rangle = 1+\theta^2 F_\phi$, 
where $F_\phi$ is the compensator F-term.
The third term in (\ref{eqn:Leff}) then becomes
$\int d^4\theta (1+\overline{\theta}^2 F_\phi^\dag)(1-\theta^2 F_\phi)\gamma_i L_i H_u$.
This suggests generation of a bilinear R-parity breaking term 
$W_{\not R}=F_\phi^\dag\gamma_i L_i H_u\equiv \mu_i L_i H_u$, as well as an R-parity breaking 
B-term $-|F_\phi|^2\gamma_i\tilde\ell_i h_u$.
In generic scenarios of supersymmetry breaking\footnote{
There are exceptional cases, e.g. the almost no-scale model \cite{Luty:2002hj}.
}
the compensator F-term gives rise to the gravitino mass, $F_\phi\approx m_{3/2}$.
Recalling that $\gamma_i=\gamma$ is related to the nonminimal coupling $\xi$ by
$\xi=\frac{\gamma}{4}-\frac 16$, 
we see $\gamma\approx 4\xi\gg 1$ when $\xi\gg 1$ and
$\gamma\sim {\C O}(1)$ when $\xi\lesssim 1$.
We will assume $\gamma\sim {\C O}(1)$ below, as one of the merits of our model
was that the extremely large nonminimal coupling that could lead to the unitarity violation
can be avoided.
The bilinear R-parity breaking term can then be written as
$
W_{\not R}\sim m_{3/2}\gamma_i L_i H_u
%W_{\not R}
\sim m_{3/2}L_k H_u.
$

In the presence of the bilinear R-parity violating terms
$W\sim H_u(\mu H_d+m_{3/2} L_k)$
the lepton number violating superpotential
\beq
W_{\Delta L=1}\sim (y_e^{ij}\varepsilon^k)e_i^cL_jL_k+(y_d^{ij}\varepsilon^k)d_i^c Q_jL_k,
\eeq
where 
$\varepsilon^k\sim m_{3/2}/\mu$,
is generated.
The cosmological constraints for these effective Yukawa couplings are 
%\cite{Dreiner:1997uz,Barbier:2004ez}
\cite{Campbell:1990fa,Fischler:1990gn,Dreiner:1992vm} %Ref. [11] of arXiv:1111.1789
\beq
y_e^{ij}\varepsilon^k, \quad 
y_d^{ij}\varepsilon^k\lesssim 10^{-7},
\label{eqn:RPB}
\eeq 
from which one obtains
$\varepsilon^k\sim m_{3/2}/\mu\lesssim 10^{-5}\cos\beta\sim10^{-6}$.
For a typical value of the MSSM $\mu$ parameter $\mu\sim 1$ TeV 
the gravitino mass of our model is
\beq
m_{3/2}\lesssim 1\mbox{ MeV}.
\label{eqn:GravitinoMass}
\eeq
Thus, supersymmetry breaking mechanisms consistent with our scenario are those giving a small 
gravitino mass (such as the gauge mediation scenario).
Note that the gravitinos with the small mass (\ref{eqn:GravitinoMass}) are still a good candidate for
the cold dark matter, even though the R-parity is broken.

The abundance of the thermally produced gravitinos is calculated as
\cite{Bolz:2000fu,Pradler:2006qh,Steffen:2006hw}
\beq%\hspace{-2mm}
\Omega_{\widetilde G} h^2\simeq 0.3\times
\left(\frac{T_{\rm RH}}{10^{10} \mbox{ GeV}}\right)
\left(\frac{100 \mbox{ GeV}}{m_{3/2}}\right)
\left(\frac{M_{\widetilde{g}}}{1\mbox{ TeV}}\right)^2,
\label{eqn:GravitinoAbundance}
\eeq
where
$h$ is the Hubble parameter in the unit of $100$ km Mpc${}^{-1}$ s${}^{-1}$
and $M_{\widetilde g}$ is the running gluino mass.
With $m_{3/2}\approx 1$ MeV and $M_{\widetilde{g}}\approx 1$ TeV, the condition for avoiding the
overdominance $\Omega_{\widetilde G} h^2\lesssim 0.1$ yields the upper bound of the reheating
temperature
$T_{\rm RH}\lesssim 10^5$ GeV.
In a previous section we estimated the reheating temperature be $T_{\rm RH}\approx 10^{5} - 10^7$ 
GeV, assuming the decay channel Higgs $\rightarrow b\bar b$.
The constraint of the gravitino abundance (\ref{eqn:GravitinoAbundance}) suggests that
$T_{\rm RH}\approx 10^5$ GeV is favoured. %a consistent reheating temperature of the scenario. 
Taking into account the Hubble expansion for the time needed for thermalisation,
$T_{\rm RH}\sim 10^5$ GeV seems to be a reasonable reheating temperature of our scenario.
To summarise, our model typically predicts a scenario of gravitino cold dark matter with the reheating
temperature $T_{\rm RH}\sim 10^5$ GeV.
The smallness of the R-parity violating terms (\ref{eqn:RPB}) guarantees that the baryon 
asymmetry generated by thermal/nonthermal leptogenesis remains without being wiped out.

We close this section with two comments. 
First, the constraint on the R-parity violation (\ref{eqn:RPB}) does not have to be taken too strict since,
for example, details of the flavour structure may slightly relax this condition
\cite{Endo:2009cv}. 
Our second comment concerns the effects on the neutrino masses.
R-parity violation induces neutrino masses without invoking right-handed neutrinos
\cite{Hall:1983id,Lee:1984kr,Lee:1984tn,Dawson:1985vr,Aulakh:1982yn,Ellis:1984gi}.
This alternative mechanism to the seesaw can be a potential threat, since if such an 
effect dominates over the usual seesaw mechanism that would give unacceptably large neutrino 
masses.
A back-of-the-envelope calculation shows that there is actually no such danger, as the constraint 
(\ref{eqn:RPB}) is more stringent than the one coming from the neutrino masses.
For example, in the bilinear R-parity breaking scenario 
\cite{Hempfling:1995wj,Nilles:1996ij,Hirsch:2000ef}
the condition on the bilinear coefficient $\mu_i$ for not generating too large neutrino masses
is $\mu_i/\mu\lesssim 10^{-3}\gg 10^{-6}$.

%
%%%%%%%%%%%%%%%%%%%%%%%%%%%%%%%%%%%%%%%%%%%%%%%%
\section{Discussion}
%%%%%%%%%%%%%%%%%%%%%%%%%%%%%%%%%%%%%%%%%%%%%%%%

We have discussed in this paper a simple model of inflationary cosmology based on the
supergravity-embedded minimal seesaw model.
The scenario is economical as it simultaneously explains various issues -- 
neutrino oscillation, the origin of the baryon asymmetry and the origin of the cold dark matter --
apart from the standard issues of big bang cosmology solved by inflation.
We have shown that the model reproduces observationally acceptable values of the cosmological
parameters, and argued that the prediction for $n_s$, $r$ of the CMB spectrum 
will be tested by satellite experiments in the near future. 
A particularly interesting feature of this model is that the seesaw mass scale is constrained by 
the CMB.
Thus far useful constraints on the (left-handed) neutrinos, such as the total neutrino masses 
$\sum m_\nu$ and the effective number of the neutrino species, have been obtained by 
observing the CMB and the large scale structure. 
However, the nature of the right-handed neutrons remains mysterious: 
being gauge singlets, their detection in colliders is virtually impossible,
nevertheless they are essential for both seesaw mechanism and leptogenesis.
In our proposal, the CMB may provide access to the physics of the right-handed neutrinos.
%The MSSM extended with the right-handed neutrinos may thus be a 
%simplest phenomenologically viable model today.

A key feature of our model is the nonminimal coupling of the D-flat direction inflaton to the
background gravitational curvature, which is naturally implemented by supergravity embedding
of the SM. 
In contrast to the nonmimimally coupled Higgs inflation type models, the coupling in our case
need not be large.
This is related to the fact that the Dirac Yukawa coupling can be very small.
Such extremely small Yukawa coupling is, nevertheless, not unnatural. 
The Dirac Yukawa coupling corresponding to a TeV scale right-handed neutrino mass in our model is in the same order as the electron Yukawa coupling. 
As Nature allows such a small coupling for the electrons, there is no reason to exclude the Dirac Yukawa coupling of the same order for the neutrinos.

Finally, we comment on extension of our model in various directions.
While embedding into the $SO(10)$ grand unified theory is not possible, 
one may for example consider our scenario in the grand unified theory of $SU(5)$ plus a singlet.
Also, type III seesaw with $SU(5)$ adjoint neutrinos in $SU(5)$ is possible.
It is also straightforward to extend our model to the right-handed neutrinos with three families.
An obvious drawback of such an extension is that the inflationary scenario will contain more
unconstrained parameters and the predictive power of the model will be reduced.

%%%%%%%%%%%%%%%%%%%%%%%%%%%%%%%%%%%%%%%%%%%%%%%%

%%%%%%%%%%%%%%%%%%%%%%%%%%%%%%%%%%%%%%%%%%%%%%%%
%{\em Acknowledgements.}---
\subsection*{Acknowledgments}
S.K. acknowledges helpful conversations with Alejandro Ibarra, Shinta Kasuya, Kazunori Kohri
and Masahide Yamaguchi.
This research was supported in part by 
the Research Program MSM6840770029 the project of International Cooperation 
ATLAS-CERN of the Ministry of Education, Youth and Sports of Czech Republic,
the JSPS - ASCR Japan - Czech Republic Research Cooperative Program (M.A.), 
the National Research Foundation of Korea
Grant-in-Aid for Scientific Research No. 2012-007575 (S.K.)
and by the DOE Grant No. DE-FG02-10ER41714 (N.O.).
A part of the numerical computation was carried out using computing facilities at the
Yukawa Institute, Kyoto University.

\bigskip

\appendix

%%%%%%%%%%%%%%%%%%%%%%%%%%%%%%%%%%%%%%%%%%%%%%%%

%%%%%%%%%%%%%%%%%%%%%%%%%%%%%%%%%%%%%%%%%%%%%%%%

\section{Boltzmann equations}
We discussed leptogenesis within our inflationary scenario in Sec. \ref{sec:lepto}
and presented the solutions of the  Boltzmann equations.
In this appendix we collect related formulae.
The Boltzmann equations in the context of leptogenesis are discussed in 
\cite{Kolb:1979qa,Dolgov:1981hv,Luty:1992un}.
We follow the conventions of \cite{Plumacher:1997ru}.

In the supersymmetric minimal seesaw model the dominant processes for 
generating the lepton number are the decay of the right-handed Majorana neutrino $N_m$
into the up-type Higgs boson and a lepton, or into the higgsino and a scalar lepton:
\beq
N_m\rightarrow H_u+\ell,\quad \overline{\tilde h}+\tilde\ell,
\eeq
as well as the decay of its superpartner $\widetilde{N_j^c}$ into the higgsino and a lepton, 
or into the Higgs boson and a scalar lepton:
\beq
\widetilde{N_m^c}\rightarrow H_u+\tilde\ell,\quad \tilde h+\overline{\ell}.
\eeq
The decay widths for these processes are
\beq
\Gamma_m=\frac{M_m}{4\pi}(y_Dy_D^\dag)_{mm}.
\label{eqn:Gamma}
\eeq
It is convenient to parametrize the inverse temperature using the right-handed neutrino mass 
$M_1$ as
\beq
z\equiv \frac{M_1}{T}.
\eeq
In the out-of-equilibrium decay of the right-handed neutrinos and sneutrinos, we assume
the initial number densities of the leptons $Y_{L_f}$ and sleptons $Y_{L_s}$ to be zero.
We also assume that the right-handed sneutrinos are initially symmetric: 
$Y_{\widetilde{N_m^c}}=Y_{\widetilde{N_m^c}^\dag}$. 
Ignoring subdominant processes, it follows %from the Boltzmann equations 
that $Y_{\widetilde{N_m^c}}=Y_{\widetilde{N_m^c}^\dag}$ persists during the subsequent evolution.
The Boltzmann equations for the number densities of the right-handed (s)neutrinos and the 
(s)lepton numbers then read 
\bea
\frac{dY_{N_m}}{dz}
&=&-\frac{z}{sH(M_1)}\left(\frac{Y_{N_m}}{Y_{N_m}^{\rm eq}}-1\right)\gamma_{N_m},
\label{eqn:boltzmann1}\\
\frac{dY_{\widetilde{N}_m^c}}{dz}
&=&-\frac{z}{sH(M_1)}\left(\frac{Y_{\widetilde{N}_m^c}}{Y_{\widetilde{N}_m^c}^{\rm eq}}
-1\right)\gamma_{N_m},\label{eqn:boltzmann2}\\
\frac{dY_{L_f}}{dz}
&=&-\frac{z}{sH(M_1)}\sum_{m}\Big\{\left(\half\frac{Y_{L_f}}{Y_{\ell}^{\rm eq}}
+\varepsilon_m\right)\gamma_{N_m}\nn\\
&&\qquad-\half\left(\frac{Y_{N_m}}{Y_{N_m}^{\rm eq}}
+\frac{Y_{\widetilde{N}_m^c}}{Y_{\widetilde{N}_m^c}^{\rm eq}}\right)
\varepsilon_m\gamma_{N_m}\Big\},
\label{eqn:boltzmann3}\\
\frac{dY_{L_s}}{dz}
&=&-\frac{z}{sH(M_1)}\sum_{m}\Big\{\left(\half\frac{Y_{L_s}}{Y_{\widetilde\ell}^{\rm eq}}
+\varepsilon_m\right)\gamma_{N_m}\nn\\
&&\qquad-\half\left(\frac{Y_{N_m}}{Y_{N_m}^{\rm eq}}
+\frac{Y_{\widetilde{N}_m^c}}{Y_{\widetilde{N}_m^c}^{\rm eq}}\right)
\varepsilon_m\gamma_{N_m}\Big\}.
\label{eqn:boltzmann4}
\eea
Here, $H(M_1)$ is the Hubble parameter at temperature $T=M_1$ and
\bea
\gamma_{N_m}=n_{N_m}^{\rm eq}\frac{K_1(z)}{K_2(z)}\Gamma_m
\eea
is the reaction density of the decay processes. 
$K_1$, $K_2$ are the elliptic integrals of the first and the second kind. 
The Boltzmann equations (\ref{eqn:boltzmann1}) and (\ref{eqn:boltzmann2}),
(\ref{eqn:boltzmann3}) and (\ref{eqn:boltzmann4}) are identical due to supersymmetry.

We solved the above equations with the yields in equilibrium,
\bea
Y_{N_m}^{\rm eq}&=&Y_{\widetilde{N}_m^c}^{\rm eq}=\frac{n_{N_m}^{\rm eq}}{s}, \quad
n_{N_m}^{\rm eq}=\frac{M_1^3}{\pi^2 z}K_2(z),\\
Y_{\ell}^{\rm eq}&=&Y_{\widetilde\ell}^{\rm eq}=\frac{n_{\ell}^{\rm eq}}{s}, \quad
n_{\ell}^{\rm eq}=\frac{2}{\pi^2}\left(\frac{M_1}{z}\right)^3.
\eea
These Boltzmann equations assume supersymmetry and are not strictly applicable 
below the supersymmetry breaking scale $T\sim$ TeV.
The deviation from the supersymmetric case is however expected to be minor as
it should naturally be within a factor of 2.

%\bea
%\gamma_{N_m}&\approx&\frac{M_1^4}{4\pi^3 x}(y_Dy_D^\dag)_{mm}K_1\left(x\right).
%\eea

%%%%%%%%%%%%%%%%%%%%%%%%%%%%%%%%%%%%%%%%%%%%%%%%

%\section{Formulae of integrals}

%%%%%%%%%%%%%%%%%%%%%%%%%%%%%%%%%%%%%%%%%%%%%%%%

%\bibliography{pheno}

%%%%%%%%%%%%%%%%%%%%%%%%%%%%%%%%%%%%%%%%%%%%%%%%
\end{document}